\documentclass[format=acmsmall, review=false, screen=true]{acmart}

\usepackage{booktabs} 
\usepackage{amsmath,amssymb,amsfonts}
\usepackage{algorithmic}
\usepackage{graphicx}
\usepackage{listings}
\usepackage{textcomp}
\def\BibTeX{{\rm B\kern-.05em{\sc i\kern-.025em b}\kern-.08em
		T\kern-.1667em\lower.7ex\hbox{E}\kern-.125emX}}

\usepackage{booktabs} 
\usepackage{multirow}
\usepackage{xspace}
\usepackage{paralist}
\usepackage{enumitem}
\usepackage{booktabs,comment,flushend,listings,multirow,tikz,todonotes}
\usepackage{graphicx,caption,subcaption}
\usepackage{amssymb}
\usepackage{ifthen}
\usetikzlibrary{arrows,calc,shapes}
\usepackage{wrapfig}
\usepackage{xcolor}
\usepackage{listings}
\usepackage{url}
\usepackage{hyperref}
\newcommand{\ie}{\textit{i.e.,}\xspace}
\newcommand{\eg}{\textit{e.g.,}\xspace}
\newcommand{\etc}{\textit{etc.}\xspace}
\newcommand{\etal}{\textit{et al.}\xspace}

\newcommand{\secref}[1]{Section~\ref{#1}\xspace}

\newcommand{\figref}[1]{Figure~\ref{#1}\xspace}

\newcommand{\tabref}[1]{Table~\ref{#1}\xspace}

\usepackage[ruled]{algorithm2e} 

\SetAlFnt{\small}
\SetAlCapFnt{\small}
\SetAlCapNameFnt{\small}
\SetAlCapHSkip{0pt}
\IncMargin{-\parindent}
	
\acmJournal{TOSEM}
\acmVolume{-}
\acmNumber{-}
\acmArticle{-}
\acmYear{2019}
\acmMonth{2}
\copyrightyear{2019}

\setcopyright{acmcopyright}
\setcopyright{acmlicensed}

\acmDOI{-.-}

\received{September 2018}

\begin{document}
	\title[Learning Bug-Fixing Patches in the Wild via Neural Machine Translation]{An Empirical Study on Learning Bug-Fixing Patches in the Wild via Neural Machine Translation}

	\author{Michele Tufano}
	\affiliation{%
	  \institution{College of William and Mary}
	  \city{Williamsburg}
	  \state{Virginia}
	  \country{USA}}
	\email{mtufano@cs.wm.edu}
	
	\author{Cody Watson}
	\affiliation{%
		\institution{College of William and Mary}
		\city{Williamsburg}
		\state{Virginia}
		\country{USA}}
	\email{cawatson@cs.wm.edu}
	
	\author{Gabriele Bavota}
	\affiliation{%
		\institution{Universit\`{a} della Svizzera italiana (USI)}
		\state{Lugano}
		\country{Switzerland}}
	\email{gabriele.bavota@usi.ch}
	
	\author{Massimiliano Di Penta}
	\affiliation{%
		\institution{University of Sannio}
		\state{Benevento}
		\country{Italy}}
	\email{dipenta@unisannio.it}
	
	\author{Martin White}
	\affiliation{%
		\institution{College of William and Mary}
		\city{Williamsburg}
		\state{Virginia}
		\country{USA}}
	\email{mgwhite@cs.wm.edu}
	
	\author{Denys Poshyvanyk}
	\affiliation{%
		\institution{College of William and Mary}
		\city{Williamsburg}
		\state{Virginia}
		\country{USA}}
	\email{denys@cs.wm.edu}

	\begin{abstract}
		Millions of open-source projects with numerous bug fixes are available in code repositories. This proliferation of software development histories can be leveraged to learn how to fix common programming bugs. To explore such a potential, we perform an empirical study to assess the feasibility of using Neural Machine Translation techniques for learning bug-fixing patches for real defects. First, we mine millions of bug-fixes from the change histories of projects hosted on GitHub, in order to extract meaningful examples of such bug-fixes. Next, we abstract the buggy and corresponding fixed code, and use them to train an Encoder-Decoder model able to translate buggy code into its fixed version. In our empirical investigation we found that such a model is able to fix thousands of unique buggy methods in the wild. Overall, this model is capable of predicting fixed patches generated by developers in 9-50\% of the cases, depending on the number of candidate patches we allow it to generate. Also, the model is able to emulate a variety of different Abstract Syntax Tree operations and generate candidate patches in a split second.
	\end{abstract}

	%
	%

	
	%
	%

	
	\maketitle
	
	\renewcommand{\shortauthors}{Tufano et al.}

	\section{Introduction}
	\label{sec:intro}

Localizing and fixing bugs is known to be an effort-prone and time-consuming task for software developers \cite{Jorgensen:2007:SRS:1248721.1248736,seacord:2003:MLS:599767,Weiss:2007:LTF:1268983.1269017}. To support programmers in this common activity, researchers have proposed a number of approaches aimed at automatically repairing programs \cite{DBLP:conf/cec/ArcuriY08,DBLP:journals/tse/GouesNFW12,LeGoues:2012,LeGoues:2012:GECCO,DBLP:conf/wcre/LeLG16,DBLP:conf/icse/KimNSK13,Long:2016:APG:2837614.2837617,Weimer:2013:LPE:3107656.3107702,Nguyen:2013:ICSE,Mechtaev:2015,Ke:2015,Xuan:2016,Mechtaev:2016, Tian:2017:ADR:3106237.3106300, Le:2017:SSS:3106237.3106309,Bhatia:2016,Pu:2016,Perkins:2009:APE:1629575.1629585, Weiss:2017:TIS:3155562.3155641, soto2018using}.
The proposed techniques either use a generate-and-validate approach, which consists of generating many repairs (\eg through Genetic Programming like GenProg~\cite{DBLP:conf/icse/WeimerNGF09,DBLP:journals/tse/GouesNFW12}), or an approach that produces a single fix~ \cite{Nguyen:2013:ICSE, Jin:2011:AAF:1993498.1993544}.
While  automated program repair techniques still face many challenges to be applied in practice, existing work has made strides to be effective in specific cases.  These approaches, given the right circumstances, substantially contribute in reducing the cost of bug-fixes for developers \cite{DBLP:conf/icse/GouesDFW12,DBLP:journals/ese/MartinezDSXM17}.

Two major problems automated repair approaches have, are producing patches acceptable for programmers and especially for generate-and-validate techniques, over-fitting patches to test cases. Qi \etal \cite{Qi:2015} found that the majority of the reported patches generated by several generate-and-validate techniques are not correct, and that such techniques mostly achieve repair by deleting pieces of functionality or by overfitting on test cases.
To cope with this problem, Le \etal \cite{DBLP:conf/wcre/LeLG16} leverages the past history of existing projects --- in terms of bug-fix patches --- and compares automatically-generated patches with existing ones.

Patches that are similar to the ones found in the past history of mined projects are considered to be more relevant. Another approach that identifies patches from past fixes is Prophet \cite{Long:2016:APG:2837614.2837617}, which after having localized the likely faulty code by running test cases, generates patches from correct code using a probabilistic model.

Our work is motivated by the following three considerations. First, automated repair approaches are based on a relatively limited and manually-crafted (with substantial effort and expertise required) set of transformations  or fixing patterns. Second, the work done by Le \etal \cite{DBLP:conf/wcre/LeLG16} shows that the past history of existing projects can be successfully leveraged to understand what a ``meaningful" program repair patch is. Third, several works have recently demonstrated the capability of advanced machine learning 
techniques, such as deep learning, to learn from relatively large software engineering datasets. Some examples of recent models that can be used in a number of software engineering tasks include: code completion \cite{Raychev:2014:CCS:2594291.2594321,White:2015}, defect prediction \cite{WangLT16}, bug localization \cite{DBLP:conf/iwpc/LamNNN17}, 
clone detection \cite{WhiteTVP16,Tufano:2018:Similarities}, code search \cite{DeepCodeSearch},   learning API sequences \cite{DBLP:conf/sigsoft/GuZZK16}, recommending method names \cite{Allamanis:2015:SAM:2786805.2786849}, learning code changes \cite{Tufano:2019:Changes} or generating Android APKs from designer's sketches \cite{Moran:TSE2018}.

Forges like GitHub provide a plethora of change history and bug-fixing commits from a large number of software projects. A machine-learning based approach can leverage this data to learn about bug-fixing activities in the wild.

In this work, we expand upon our original idea of learning bug-fixes \cite{Tufano:2018:EIL:3238147.3240732} and extensively evaluate the suitability of a Neural-Machine Translation (NMT-based approach) to automatically generate patches for buggy code.

Automatically learning from bug-fixes in the wild provides the ability to emulate real patches written by developers. Additionally, we harness the power of NMT to ``translate'' buggy code into fixed code thereby emulating the combination of Abstract Syntax Tree (AST) operations performed in the developer written patches.
Further benefits include the static nature of NMT when identifying candidate patches, since, unlike some generate-and-validate approaches,  we do not need to execute test cases during patch generation\cite{Yang:2017:BTC:3106237.3106274, Smith:2015:CWD:2786805.2786825}. Test case execution on the  patches recommended by the NMT approach would still be necessary in  practice, however, this would only be needed on the candidate set of patches.

To this aim, we first mine a large set of ($\sim787k$) bug-fixing commits from GitHub. From these commits, we extract method-level AST edit operations using fine-grained source code differencing \cite{DBLP:conf/kbse/FalleriMBMM14}. We identify multiple method-level differences per bug-fixing commit and independently consider each one, yielding to $\sim2.3M$ bug-fix pairs (BFPs). After that, the code of the BFPs is abstracted to make it more suitable for the NMT model. Finally, an encoder-decoder model is used to understand how the buggy code is transformed into fixed code. Once the model has been trained, it is used to generate patches for unseen code.

We empirically investigate the potential of NMT to generate candidate patches that are identical to the ones implemented by developers. Also, we quantitatively and qualitatively analyze the AST operations the NMT model is able to emulate when fixing bugs. Finally, we evaluate its efficiency by computing the time needed to learn a model and to infer patches.

The results indicate that trained NMT models are able to successfully predict the fixed code, given the buggy code, in 9-50\% of the cases. The percentage of bugs that can be fixed depends on the number of candidate patches the model is required to produce. We find that over 82\% of the generated candidate patches are syntactically correct. When performing the \textit{translation} the models emulate between 27-64\% of the AST operations used by developers to fix the bugs, during patch generation. The NMT models are capable of producing multiple candidate patches for a given buggy code in less then a second.

In all, the paper provides the following contributions:

\begin{itemize}
	\item An extensive empirical investigation into the applicability of NMT techniques for learning how to generate patches from bug-fixes;
	\item A detailed process for training and evaluating NMT models by mining, extracting, and abstracting bug-fixing examples in the wild;
	\item A publicly available replication package, including datasets, source code, tools, and detailed results reported and discussed in this study \cite{online-appendix}.
\end{itemize}

	\section{Approach}
	\label{sec:design}

\begin{figure*}[t!]
	\centering
	\includegraphics[width=\linewidth]{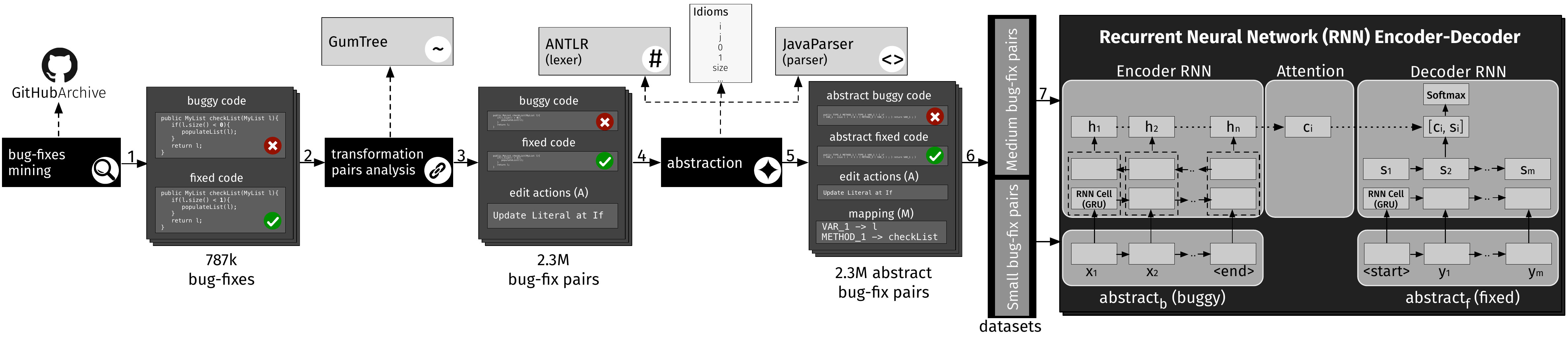}
	\caption{Overview of the process used to experiment with an NMT-based approach.}
	\label{fig:approach}
\end{figure*}

Fig. \ref{fig:approach} shows an overview of the NMT approach that we experiment with. The dark boxes represent the main phases, the arrows indicate data flows, and the dashed arrows denote dependencies on external tools or data. We mine bug-fixing commits from thousands of GitHub repositories using GitHub Archive \cite{githubarchive} (Section \ref{sec:mining}). From the bug-fixes, we extract method-level pairs of \textit{buggy} and corresponding \textit{fixed} code named \textit{bug-fix pairs} (BFPs) (Section \ref{sec:tp_extraction}). BFPs are the examples that we use to learn how to fix code from bug-fixes (\textit{buggy} $\rightarrow$ $fixed$). We use GumTree \cite{DBLP:conf/kbse/FalleriMBMM14} to identify the list of edit actions ($A$) performed between the buggy and fixed code. Then, we use a Java Lexer and Parser to abstract the source code of the BFPs (Section \ref{sec:tp_abstraction}) into a representation better suited for learning. During the abstraction, we keep frequent identifiers and literals we call \textit{idioms} 
within the representation. The output of this phase are the abstracted BFPs and their corresponding mapping $M$, which allows reconstructing the original source code. Next, we generate two datasets of BFPs grouping together fixes for small and medium methods, respectively (Section \ref{sec:filtering}). Finally, for each set, we use an encoder-decoder model to learn how to transform a \textit{buggy} code into a corresponding \textit{fixed} version (Section \ref{sec:learning}). The trained models can be used to generate patches for unseen buggy code.

\subsection{Bug-Fixes Mining}
\label{sec:mining}
We downloaded from GitHub Archive \cite{githubarchive} every public GitHub event between March 2011 and October 2017 and we used the Google BigQuery APIs to identify all commits having a message containing the patterns \cite{DBLP:conf/icsm/FischerPG03}:  (``fix'' or ``solve'') and (``bug'' or ``issue'' or ``problem'' or ``error'').  We identified $\mathtt{\sim}$10M (10,056,052) bug-fixing commits. 

As the content of commit messages and issue trackers might imprecisely identify bug-fixing commits \cite{AntoniolAPKG08, HerzigJZ13}, two authors independently analyzed a statistically significant sample (95\% confidence level $\pm5\%$ confidence interval, for a total size of 384) of identified commits to check whether they were actually bug fixes. After solving 13 cases of disagreement, they concluded that 97.6\% of the identified bug-fixing commits were true positive. Details about this evaluation are in our online appendix \cite{online-appendix}. 

For each bug-fixing commit, we extracted the source code before and after the bug-fix using the GitHub Compare API \cite{github-compare}. This allowed us to collect the buggy (pre-commit) and the fixed (post-commit) code. We discarded commits related to non-Java files, as well as files that were created in the bug-fixing commit, since there would be no buggy version to learn from. Moreover, we discarded commits impacting more than five Java files, since we aim to learn focused bug-fixes that are not spread across the system.

The result of this process was the buggy and fixed code of 787,178 bug-fixing commits.

\subsection{Bug-Fix Pairs Analysis}
\label{sec:tp_analysis}
A BFP (Bug-Fixing Pair) is a pair $(m_b, m_f)$ where $m_b$ represents a buggy code component and $m_f$ represents the corresponding fixed code. We will use these BFPs to train the NMT model, make it learning the translation from buggy ($m_b$) to fixed ($m_f$)  code, thus being able of generating patches.

\subsubsection{Extraction}
\label{sec:tp_extraction}
Given $(f_b, f_f)$ a pair of buggy and fixed file from a bug-fix \textit{bf}, we used the GumTree Spoon AST Diff tool \cite{DBLP:conf/kbse/FalleriMBMM14} to compute the AST differencing between $f_b$ and $f_f$. This computes the sequence of edit actions performed at the AST level that allows to transform the $f_b$'s AST into the $f_f$'s AST.

GumTree Diff considers the following edit actions: (i) \textit{updatedValue}: replaces the value of a node in the AST (\texttt{<Update, AST\_Node\_Type, Target\_AST\_Node\_Type>}); (ii)  \textit{add/insert}: inserts a new node in the AST (\texttt{<Insert, AST\_Node\_Type, Target\_AST\_Node\_Type>}); (iii) \textit{delete}: which deletes a node in the AST (\texttt{<Delete, AST\_Node\_Type, Target\_AST\_Node\_Type>}); (iv) \textit{move}: moves an existing node in a different location in the AST (\texttt{<Move, AST\_Node\_Type, Source\_AST\_Node\_Type, Target\_AST\_Node\_Type>}). In our analysis we consider the set of AST edit actions as defined by GumTree Diff.

Since the file-level granularity could be too large to learn patterns of transformation, we separate the code into method-level fragments that will constitute our BFPs. The rationale for choosing method-level BFPs is supported by several reasons. First, methods represent a reasonable target for fixing activities, since they are likely to implement a single task or functionality. Second, methods provide enough meaningful context for learning fixes, such as variables, parameters, and method calls used in the method. This choice is justified by recent empirical studies, which indicated how the large majority of fixing patches consist of single line, single churn or, worst cases, churns separated by a single line \cite{DBLP:conf/wcre/SobreiraDDMM18}. 

Smaller snippets of code lack the necessary context and, hence, they could not be considered. Finally, considering arbitrarily long snippets of code, such as hunks in diffs, makes learning more difficult given the variability in size and context \cite{10.1007/978-3-642-35843-2_6,DBLP:conf/iwpc/AlaliKM08}. 

We first rely on GumTree to establish the mapping between the nodes of $f_b$ and $f_f$. Then, we extract the list of mapped pairs of methods $L = \{(m_{1b}, m_{1f}), \dots, (m_{nb}, m_{nf}) \}$. Each pair $(m_{ib}, m_{if})$ contains the method $m_{ib}$ (from the buggy file $f_b$) and the corresponding method $m_if$ (from the fixed file $f_f$). Next, for each pair of mapped methods, we extract a sequence of edit actions using the GumTree algorithm. We then consider only those method pairs for which there is at least one edit action (\ie we disregard methods that have not been modified during the fix). Therefore, the output of this phase is a list of $BFPs = \{bfp_1, \dots, bfp_k\}$, where each BFP is a triplet $bfp = \{m_b, m_f, A\}$, where $m_b$ is the buggy method, $m_f$ is the corresponding fixed method, and $A$ is a sequence of edit actions that transforms $m_b$ in $m_f$. We exclude methods created/deleted during the fixing, since we cannot learn fixing operations from them. Overall, we extracted $\sim$2.3M BFPs. 

It should be noted that the process we use to extract the BFPs: (i) does not capture changes performed outside methods (\eg class signature, attributes, \etc), and (ii) considers each BFP as an independent bug fix, meaning that multiple methods modified in the same bug fixing activity are considered independently from one another.

\subsubsection{Abstraction}
\label{sec:tp_abstraction}
Learning bug-fixing patterns is extremely challenging by working at the level of raw source code. This is especially due to the huge vocabulary of terms used in the identifiers and literals of the $\mathtt{\sim}$2M mined projects. Such a large vocabulary would hinder our goal of learning transformations of code as a NMT task. For this reason, we abstract the code and generate an expressive yet vocabulary-limited representation. We use a Java lexer and a parser to represent each buggy and fixed method within a BFP as a stream of tokens. The lexer, built on top of ANTLR \cite{Parr:2011:LFA:1993498.1993548, Parr:2013:DAR:2501720}, tokenizes the raw code into a stream of tokens, that is then fed into a Java parser \cite{javaparser}, which discerns the role of each identifier (\ie whether it represents a variable, method, or type name) and the type of a literal. 

Each BFP is abstracted in isolation. Given a BFP $bfp=\{m_b, m_f, A\}$, we first consider the source code of $m_b$. The source code is fed to a Java lexer, producing the stream of tokens. The stream of tokens is then fed to a Java parser, which recognizes the identifiers and literals in the stream. The parser generates and substitutes a unique ID for each identifier/literal within the tokenized stream. If an identifier or literal appears multiple times in the stream, it will be replaced with the same ID. The mapping of identifiers/literals with their corresponding IDs is saved in a map ($M$). The final output of the Java parser is the abstracted method ($abstract_b$). Then, we consider the source code of $m_f$. The Java lexer produces a stream of tokens, which is then fed to the parser. The parser continues to use a map $M$ when abstracting $m_f$. The parser generates new IDs only for novel identifiers/literals, not already contained in $M$, meaning, they exist in $m_f$ but not in $m_b$. Then, it replaces all the identifiers/literals with the corresponding IDs, generating the abstracted method ($abstract_f$). The abstracted BFP is now a 4-tuple $bfpa = \{abstract_b, abstract_f, A, M\}$, where $M$ is the ID mapping for that particular BFP. The process continues considering the next BFP, generating  a  new mapping $M$. Note that we first analyze the buggy code $m_b$ and then the corresponding fixed code $m_f$ of a BFP, since this is the direction of the learning process.

IDs are assigned to identifiers and literals in a sequential and positional fashion: The first method name found will be assigned the ID of \texttt{METHOD\_1}, likewise the second method name will receive the ID of \texttt{METHOD\_2}. This process continues for all the method and variable names (\texttt{VAR\_X}) as well as the literals (\texttt{STRING\_X}, \texttt{INT\_X}, \texttt{FLOAT\_X}). 

At this point, $abstract_b$ and $abstract_f$ of a BFP are a stream of tokens consisting of language keywords (\eg \texttt{for}, \texttt{if}), separators (\eg ``('', ``;'', ``\}'') and IDs representing identifiers and literals. Comments and annotations have been removed from the code representation.

Some identifiers and literals appear so often in the code that, for the purpose of our abstraction, they can almost be treated as keywords of the language. This is the case for the variables \texttt{i}, \texttt{j}, or \texttt{index}, that are often used in loops, or for  literals such as \texttt{0}, \texttt{1}, \texttt{-1}, often used in conditional statements and return values. Similarly, method names, such as \texttt{size} or \texttt{add}, appear several times in our code base, since they represent common concepts. These identifiers and literals are often referred to as ``idioms'' \cite{Brown:2017:CFW:3106237.3106280}.  We include idioms in our representation and do not replace idioms with a generated ID, but rather keep the original text when abstracting the code. 

To define the list of idioms, we first randomly sampled 300k BFPs and considered all their original source code. Then, we extracted the frequency of each identifier/literal used in the code, discarding keywords, separators, and comments. Next, we analyzed the distribution of the frequencies and focused on the top $0.005\%$ frequent words (outliers of the distribution).  Two authors manually analyzed this list and curated a set of 272 idioms also including standard Java types such as \texttt{String}, \texttt{Integer}, common \texttt{Exceptions}, \etc The list of idioms is available in the online appendix \cite{online-appendix}.

This representation provides enough context and information to effectively learn code transformations, while keeping a limited vocabulary ($|V| = \mathtt{\sim}$430). The abstracted code can be mapped back to the real source code using the mapping ($M$).

\begin{figure}[h]
	\centering
	\includegraphics[width=0.9\linewidth]{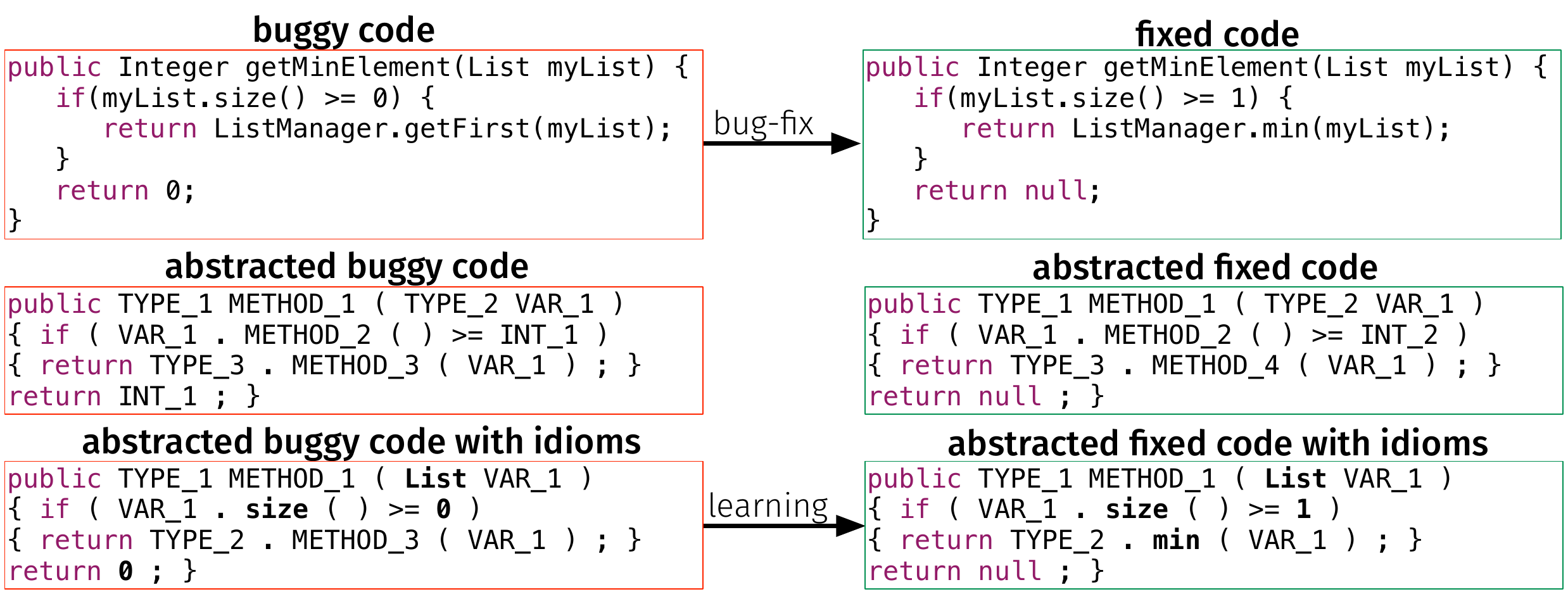}
	\caption{Code Abstraction Example.}
	\label{fig:example}
\end{figure}

To better understand our representation, let us consider the example in Fig. \ref{fig:example}, where we see a bug-fix related to finding the minimum value in a list of integers. 
The buggy method contains three errors, which the fixed code rectifies. The first bug is within the if-condition, where the buggy method checks if the list size is greater than or equal to \texttt{0}. This is problematic since a list without any values cannot have a minimum value to return. The second bug is in the method called \texttt{getFirst}, this will return the first element in the list, which may or may not be the minimum value. Lastly, if the if-condition fails in the buggy method then the method returns \texttt{0}; returning \texttt{0} when the minimum is unable to be identified is incorrect as it indicates that one of the elements within the list is \texttt{0}. The fixed code changes the if-condition to compare against a list size of \texttt{1} rather than \texttt{0}, uses the \texttt{min} method to return the minimum value and changes the return value to \texttt{null} when the if-condition fails. 

Using the buggy and fixed code for training, although a viable and realistic bug-fix, presents some issues. When we feed the buggy piece of code to the Java Parser and Lexer, we identify some problems with the mapping. For example, the abstracted fixed code contains \texttt{INT\_2} and \texttt{METHOD\_4}, which are not contained in the abstracted version of the buggy code or its mapping. Since the mapping of tokens to code is solely reliant on the buggy method, this example would require the synthesis of new values for \texttt{INT\_2} and \texttt{METHOD\_4}. However, the methodology takes advantage of idioms, allowing to still consider this BFP. When using the abstraction with idioms, we are able to replace tokens with the values they represent. Now, when looking at the abstracted code with idioms for both buggy and fixed code, there are no abstract tokens found in the fixed code that are not in the buggy code. Previously, we needed to synthesize values for \texttt{INT\_2} and \texttt{METHOD\_4}, however, \texttt{INT\_2} was replaced with idiom \texttt{1} and \texttt{METHOD\_4} with idiom \texttt{min}. With the use of idioms, we are capable of keeping this BFP while maintaining the integrity of learning real, developer-inspired patches.

\subsubsection{Filtering}
\label{sec:filtering}
\begin{figure}[h]
	\centering
	\includegraphics[width=0.55\linewidth]{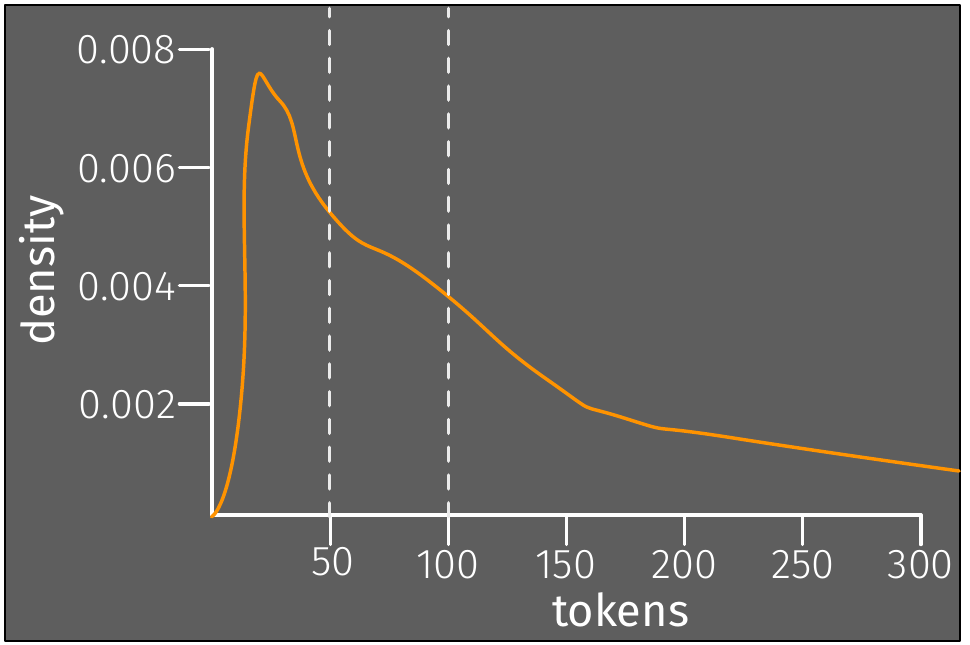}
	\caption{Distribution of BFPs by the number of tokens.}
	\label{fig:density}
\end{figure}

We filter out BFPs that: (i) contain lexical or syntactic errors (\ie either the lexer or parser fails to process them) in either the buggy or fixed code; (ii) their buggy and fixed abstracted code ($abstract_b$, $abstract_f$) resulted in equal strings; (iii) performed more than 100 atomic AST actions ($|A|>100$) between the buggy and fixed version.
The rationale behind the latter decision was to eliminate outliers of the distribution (the 3rd quartile of the distribution is 14 actions), which could hinder the learning process. Moreover, we do not aim to learn such large bug-fixing patches. Next, we analyze the distribution of BFPs based on their size, measured in the number of tokens, shown in Fig. \ref{fig:density}. We can notice that the density of BFPs for the buggy code has a peak before 50 tokens and a long tail that extends over 300 tokens. 

NMT models require large training dataset in order to achieve reasonable results. Moreover, the variability in sentences length can affect training and performance of the models, even when techniques such as bucketing and padding are employed. For these reasons, we decided to focus on the intervals where most of the data points are available. From Fig. \ref{fig:density} it is clear that most of the data points are concentrated in the interval 0-100. Further analysis showed that there are more data points in the interval 0-100 than in the larger interval 100-500. Therefore, we disregard long methods (longer than 100 tokens) and focused on small/medium size BFPs. We create two datasets: $BFP_{small} = \{ bfp \leq 50 \}$ and $BFP_{medium} = \{ 50 < bfp \leq 100 \}$. 

\subsubsection{Synthesis of Identifiers and Literals}

BFPs are the examples we use to make our model learn how to fix source code. Given a $bfp = \{m_b, m_f, A\}$, we first abstract its code, obtaining $bfpa = \{abstract_b, abstract_f, A, M\}$. The buggy code $abstract_b$ is used as input to the model, which is trained to output the corresponding fixed code $abstract_f$. This output can then be mapped back to real source code using $M$.

In the real usage scenario, when the model is deployed, we do not have access to the oracle (\ie fixed code, $abstract_f$), but only to the input code. This source code can then be abstracted and fed to the model, which generates as output a predicted code ($abstract_p$). The IDs that the $abstract_p$ contains can be mapped back to real values only if they also appear in the input code. If the fixed code suggests to introduce a method call, \texttt{METHOD\_6}, which is not found in the input code, we cannot automatically map \texttt{METHOD\_6} to an actual method name. This inability to map back source code exists for any newly created ID generated for identifiers or literals, which are absent in the input code.

Therefore, it appears that the abstraction process, which allows us to limit the vocabulary size and facilitate the training process, confines us to only learning fixes that re-arrange keywords, identifiers, and literals already available in the context of the buggy method. This is the primary reason we decided to incorporate idioms in our code representation, and treat them as keywords of the language. Idioms help  retaining BFPs that otherwise would be discarded because of the inability to synthesize new identifiers or literals. This allows the model to learn how to replace an abstract identifier/literal with an idiom or an idiom with another idiom (\eg bottom part of Fig. \ref{fig:example}).

After these filtering phases, the two datasets $BFP_{small}$ and $BFP_{medium}$ consist of 58k (58,350) and 65k (65,455) bug-fixes, respectively.

\subsection{Learning Patches}
\label{sec:learning}
\subsubsection{Dataset Preparation}
Given a set of BFPs (\ie $BFP_{small}$ and  $BFP_{medium}$) we use the instances to train an Encoder-Decoder model. Given a $bfpa = \{abstract_b, abstract_f, A, M\}$ we use only the pair ($abstract_b, abstract_f$) of buggy and fixed abstracted code for learning. No additional information about the possible fixing actions ($A$) is provided during the learning process to the model. The given set of BFPs is randomly partitioned into: training (80\%), validation (10\%), and test (10\%) sets. Before the partitioning, we make sure to remove any duplicated pairs ($abstract_f, abstract_b$) to not bias the results, \ie same pair both in training and test set. Specifically, code duplication represents a significant threat to machine learning approaches on source code which could lead to inflated results, as recently described by Allamanis \cite{Allamanis:duplication}. In particular, there exists no pair $<bfp_i, bfp_j>$ in our dataset (\ie train, validation, and test combined) such that $bfp_i = bfp_j$. The duplicates are removed after the abstraction process, allowing us not only to discard instances with identical source code, but also those with similar code that becomes identical after the abstraction. As a matter of fact, by replacing identifiers and literals with IDs such as \texttt{VAR} and \texttt{TYPE}, we perform a process similar to the one applied by clone detection tools to identify similar code. Therefore, we go beyond simply removing instances having the same source code.

In order to better check against the presence of clones, we also employ a state-of-the-art clone detection tool, NiCad \cite{DBLP:conf/iwpc/RoyC08a}. 
NiCad provides pre-defined configurations for different types of clones, which allow us to avoid to set an arbitrary similarity threshold.
Also, recent work \cite{Sajnani:2016:SSC:2884781.2884877} showed the superiority of NiCad to alternative tools in terms of precision/recall. The only comparable tool is SourcererCC \cite{Sajnani:2016:SSC:2884781.2884877}, which is above all better in terms of scalability, something not really relevant for our application scenario.

We found no Type I clones, and only $\sim$1,200 and $\sim$400 Type II clone pairs in $BFP_{small}$ and  $BFP_{medium}$, respectively. These few Type II clones come from the idiomatic abstraction process, where instances differ for idioms in the abstract code. Note that, even if these instances represent clones, they are effectively two different inputs/outputs for the model, since their sequences of tokens are different.

\subsubsection{NMT}
The experimented models are based on an RNN Encoder-Decoder architecture, commonly adopted in NMT \cite{kalchbrenner-blunsom:2013:EMNLP, DBLP:journals/corr/SutskeverVL14, DBLP:journals/corr/ChoMGBSB14}.  This model consists of two major components: an RNN Encoder, which \textit{encodes} a sequence of terms \boldmath{$x$} into a vector representation, and an RNN Decoder, which \textit{decodes} the representation into another sequence of terms $y$. The model learns a conditional distribution over a (output) sequence conditioned on another (input) sequence of terms: \unboldmath$P(y_1,.., y_m | x_1,.., x_n)$, where $n$ and $m$ may differ. In our case, given an input sequence $\mathbf{x} = abstract_b = (x_1,.., x_n)$ and a target sequence $\mathbf{y} = abstract_f = (y_1,.., y_m)$, the model is trained to learn the conditional distribution:  $P(abstract_f | abstract_b) = P(y_1,.., y_m | x_1,.., x_n)$, where $x_i$ and $y_j$ are abstracted source tokens: Java keywords, separators, IDs, and idioms. Fig. \ref{fig:approach} shows the  architecture of the Encoder-Decoder model with attention mechanism \cite{DBLP:journals/corr/BahdanauCB14, DBLP:journals/corr/LuongPM15, DBLP:journals/corr/BritzGLL17}. The Encoder takes as input a sequence $\mathbf{x} = (x_1,.., x_n)$ and produces a sequence of states $\mathbf{h} = (h_1,.., h_n)$. We rely on a bi-directional RNN Encoder \cite{DBLP:journals/corr/BahdanauCB14}, which is formed by a backward and a forward RNN, which are able to create representations taking into account both past and future inputs \cite{DBLP:journals/corr/BritzGLL17}. That is, each state $h_i$ represents the concatenation (dashed box in Fig. \ref{fig:approach}) of the states produced by the two RNNs when reading the sequence in a forward and backward fashion: $h_i = [ \overrightarrow{h_i}; \overleftarrow{h_i}] $. 

The RNN Decoder predicts the probability of a target sequence $\mathbf{y} = (y_1,.., y_m)$ given $\mathbf{h}$. Specifically, the probability of each output term $y_i$ is computed based on: (i) the recurrent state $s_i$ in the Decoder; (ii) the previous $i-1$ terms $(y_1,.., y_{i-1})$; and (iii) a context vector $c_i$. The latter constitutes the attention mechanism. The vector $c_i$ is computed as a weighted average of the states in $\mathbf{h}$: $c_i = \sum_{t=1}^{n}{a_{it}h_t}$ where the weights $a_{it}$ allow the model to pay more \textit{attention} to different parts of the input sequence. Specifically, the weight $a_{it}$ defines how much the term $x_i$ should be taken into account when predicting the target term $y_t$.

The entire model is trained end-to-end (Encoder and Decoder jointly) by minimizing the negative log likelihood of the target terms, using stochastic gradient descent.

\subsubsection{Generating Multiple Patches via Beam Search}

\begin{figure*}[t!]
	\centering
	\includegraphics[width=1\linewidth]{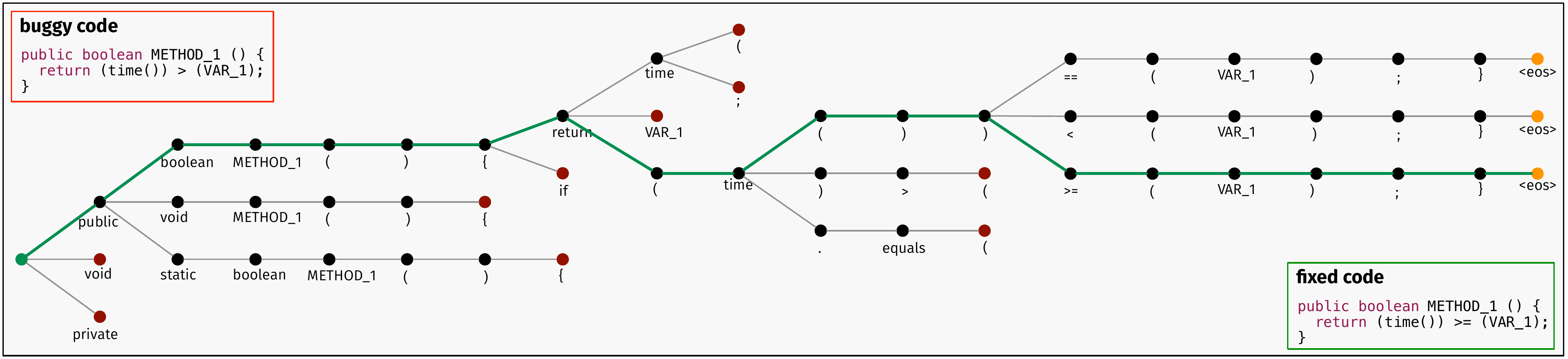}
	\caption{Beam Search Visualization.}
	\label{fig:beamsearch}
\end{figure*}

After the model is trained, it is evaluated against the test set of \textit{unseen} buggy code. The classic greedy decoding selects, at each time step $i$, the output term $y_i$ with the highest probability. The downside of this decoding strategy is that, given a buggy code as input, the trained model will generate only one possible sequence of predicted fixed code. Conversely, we would like to generate multiple potential patches (\ie sequence of terms representing the fixed code) for a given buggy code. To this aim, we employ a different decoding strategy called Beam Search and used in previous applications of deep learning \cite{DBLP:journals/corr/abs-1211-3711, boulanger2013audio, DBLP:journals/corr/BahdanauCB14, Raychev:2014:CCS:2594291.2594321}.

The major intuition behind Beam Search decoding is that rather than predicting at each time step the token with the best probability, the decoding process keeps track of $k$ hypotheses (with $k$ being the beam size or width). Formally, let $\mathcal{H}_t$ be the set of $k$ hypotheses decoded till time step $t$:

$$
\mathcal{H}_t = \{ (\tilde{y}^{1}_1, \dots , \tilde{y}^{1}_t) , (\tilde{y}^{2}_1, \dots , \tilde{y}^{2}_t) , \dots , (\tilde{y}^{k}_1, \dots , \tilde{y}^{k}_t)   \}
$$

At the next time step $t+1$, for each hypothesis there will be $|V|$ possible $y_{t+1}$ terms ($V$ being the vocabulary), for a total of $k \cdot |V|$ possible hypotheses.

$$
\mathcal{C}_{t+1} = \bigcup\limits_{i=1}^{k}  \{ (\tilde{y}^{i}_1, \dots , \tilde{y}^{i}_t, v_1) , \dots , (\tilde{y}^{i}_1, \dots , \tilde{y}^{i}_t, v_{|V|})   \}
$$

From these candidate sets, the decoding process keeps the $k$ sequences with the highest probability. The process continues until each hypothesis reaches the special token representing the end of a sequence. We consider these $k$ final sentences as candidate patches for the buggy code. Note that when $k=1$, Beam Search decoding coincides with the greedy strategy.

Fig. \ref{fig:beamsearch} shows an example of the Beam Search decoding strategy with $k=3$.  Given the $abstract_b$ code as input (top-left), the Beam Search starts by generating the top-3 candidates for the first term (\ie \texttt{public}, \texttt{void}, \texttt{private}). At the next time step, the beam search expands each current hyphothesis and finds that the top-3 most likely are those following the node \texttt{public}. Therefore, the other two branches (\ie \texttt{void}, \texttt{private}) are pruned (\ie red nodes). The search continues till each hypothesis reaches the \texttt{<eos>} (End Of Sequence) symbol. Note that each hypothesis could reach the end at different time steps. This is a real example generated by our model, where one of the candidate patches is the actual fixed code (\ie green path).

\subsubsection{Hyperparameter Search}
\label{sec:hyperpar}
For both models built on the $BFP_{small}$ and $BFP_{medium}$ dataset (\ie $M_{small}$ and $M_{medium}$) we performed hyperparameter search by testing ten configurations of the encoder-decoder architecture. The configurations tested different combinations of RNN Cells (LSTM \cite{Hochreiter:1997:LSM:1246443.1246450} and GRU \cite{DBLP:journals/corr/ChoMGBSB14}), number of layers (1, 2, 4) and units (256, 512) for the encoder/decoder, and the embedding size (256, 512). Bucketing and padding was used to deal with the variable length of the sequences. We trained our models for a maximum of 60k epochs, and selected the model's checkpoint before over-fitting the training data. To guide the selection of the best configuration, we used the loss function computed on the \textit{validation} set (not on the test set), while the results  are computed on the \textit{test} set. All the configurations and settings are available in our online appendix \cite{online-appendix}.

\subsubsection{Code Concretization}
\label{sec:concretization}
In this final phase, the abstracted code generated as output by the NMT model is concretized by mapping back all the identifiers and literal IDs to their actual values. The process simply replaces each ID found in the abstracted code to the real identifier/literal associated with the ID and saved in the mapping $M$, for each method pair. The code is automatically indented and additional code style rules can be enforced during this stage. While we do not deal with comments, they could be reintroduced in this stage as well.

	\section{Experimental Design}
	\label{sec:exp_design}
The {\em goal} of this study is, as stated in the introduction, to empirically assess whether NMT can be used to learn fixes in the wild. The {\em context} consists of a dataset of  bug fixes mined from Java open source projects hosted on GitHub (see Section \secref{sec:design}).

The study aims at answering three research questions, described in the following.

\subsection{RQ1: Is Neural Machine Translation a viable approach to learn how to fix code?}
We aim to empirically assessing whether NMT is a viable approach to learn transformations of the code from a buggy to a fixed state. To this end, we rely on multiple internal and external datasets.

\subsubsection{Internal Dataset}
We use the two datasets $BFP_{small}$ and $BFP_{medium}$ to train and evaluate two NMT models $M_{small}$ and $M_{medium}$. Precisely, given a BFP dataset, we train different configurations of the Encoder-Decoder models, then select the best performing configuration on the validation set. We then evaluate the validity of the model with the unseen instances of the test set.

The evaluation is performed as follows: let $M$ be a trained model ($M_{small}$ or $M_{medium}$) and $T$ be the test set of BFPs ($BFP_{small}$ or $BFP_{medium}$), we evaluate the model $M$ for each $bfp = (abstract_b, abstract_f) \in T$. Specifically, we feed the buggy code $abstract_b$ to the model $M$, performing inference with Beam Search Decoding for a given beam size $k$. The model will generate $k$ different potential patches $P = \{abstract_p^1, \dots , abstract_p^k\}$. We say that the model generated a successful fix for the code if there exists an $abstract_p^i \in P$ such that $abstract_p^i = abstract_f$. We report the raw count and percentage of successfully fixed BFPs in the test set, varying the beam size $k$ from 1 (\ie a single patch is created by $M$) to 50 (\ie 50 patches are created) with incremental steps of 5.

\subsubsection{External Dataset}
The external dataset comprises methods extracted from CodRep: a Machine Learning on Source Code Competition \cite{arXiv-1807.03200}. The goal of the competition is to predict where to insert a specific line into a source code file. The competition aims at being a common playground on which the machine learning and the software engineering research communities can interact \cite{arXiv-1807.03200}, with potential usages in the field of automated program repair \cite{codrep}. The dataset is composed of five collections of source code files taken from real commits in open-source projects, belonging to seven different studies in the literature focusing on bug-fixes and change history analysis \cite{Zhong:2015:ESR:2818754.2818864, monperrus:hal-00769121, DBLP:journals/corr/Li0KBLT16, Scholtes2016, Zhou:2012:BFM:2337223.2337226, doi:10.1002/smr.1797, ICSE.2012.6227193}. On the $CodRep$ dataset we perform the same steps described in \secref{sec:tp_analysis}, which involves extracting the changed methods, abstracting the pairs, and selecting only small and medium methods. \tabref{tab:codrep} reports the number of unique (\ie no duplicates) method pairs in each of the five collections as well as the entire $CodRep$ dataset (\ie last row in \tabref{tab:codrep}) where also inter-collections duplicates have been removed. The same models $M_{small}$ and $M_{medium}$, which have been trained on the $BFP$ dataset, are then evaluated on the entire $CodRep$ dataset, without making any adjustments or adding more idioms. This helps in assessing the ``portability'' of our models when trained on a dataset $D_i$ and tested on a different dataset $D_j$.

\begin{table}[t]
	\scriptsize
	\caption{$CodRep$ Datasets.}
	\label{tab:codrep}
	\centering
	\begin{tabular}{l|c|c|}
		\toprule
		Datasets & $M_{small}$ & $M_{medium}$\\
		\midrule
		CodRep1 & 221 & 503\\
		CodRep2 & 530 & 1130\\
		CodRep3 & 665 & 1397\\
		CodRep4 & 603 & 1124\\
		CodRep5 & 1066 & 2119\\
		\midrule
		$CodRep$ & 3027 & 6205\\
		\bottomrule
	\end{tabular}
\end{table}

\subsection{RQ2: What types of operations are performed by the models?}
This RQ investigates the quality and type of operations performed by the fixes that our model generates. We perform the investigation by means of automated and manual analysis.

We first analyze the syntactic correctness of the patches for all beam widths. That is, we feed each potential patch $abstract_p^i$ to a Java lexer and parser in order to assess whether the patch is lexically and syntactically correct. We do not assess the compilability of the patches, since it would require us to download the exact, entire snapshot of each GitHub project. This would entail downloading thousands of different GitHub projects and attempting to compile them with the newly generated patch. There are also obstacles when dealing with different building systems.

Next, we focus on the BFPs that are successfully fixed by the models and analyze the types of AST operations performed during the fix. While these NMT models do not technically operate on the source code's AST, but rather on  sequences of tokens, it is still worthwhile to understand the types of AST operations that such models can emulate. This analysis will provide an idea on the potential and/or limitations of such models. In detail, we extract the AST operations by selecting the action set $A$ of the BFPs successfully fixed by the model. We identify the set $M_A$ of unique AST actions performed by the model $M$ in the successful fixes and compare them with the overall set $O_A$ of unique AST operations contained within the entire test set of BFPs (\ie those that are needed to fix all the bugs in our test sets). With this information we can compute the percentage of AST actions in $O_A$ that are learned and applied by $M$ (\ie $|M_A|/|O_A|$). We also calculate the ``theoretical bug coverage'' ensured by $M_A$ as the percentage of bugs in the test set that could be theoretically fixed by only using a subset of operations in $M_A$. This allows us to check whether the AST operations that are not ``learned'' by $M$ (\ie $|O_A| \setminus|M_A|$) are used in many bug-fixing activities, thus representing an important loss for our model. A low theoretical bug coverage indicates that many bugs in test sets can not be fixed by only using the operations in $M_A$, while a high theoretical bug coverage points to the fact that the operations not learned by $M$ are only sporadically used to fix bugs.

Finally, we discuss some interesting examples of the patches generated by NMT models.

\subsection{RQ3: What is the training and inference time of the models?}
In this RQ we evaluate the  performance of the models in terms of execution time. Specifically, we analyze and discuss the time required to train the models, and the time needed to perform an inference once models have been deployed. For the latter, we report the total time of inference and compute the average time per patch generated for every beam width.
	
	\section{Results}
	\label{sec:results}

\begin{figure*}[t!]
	\centering
	\includegraphics[width=1\linewidth]{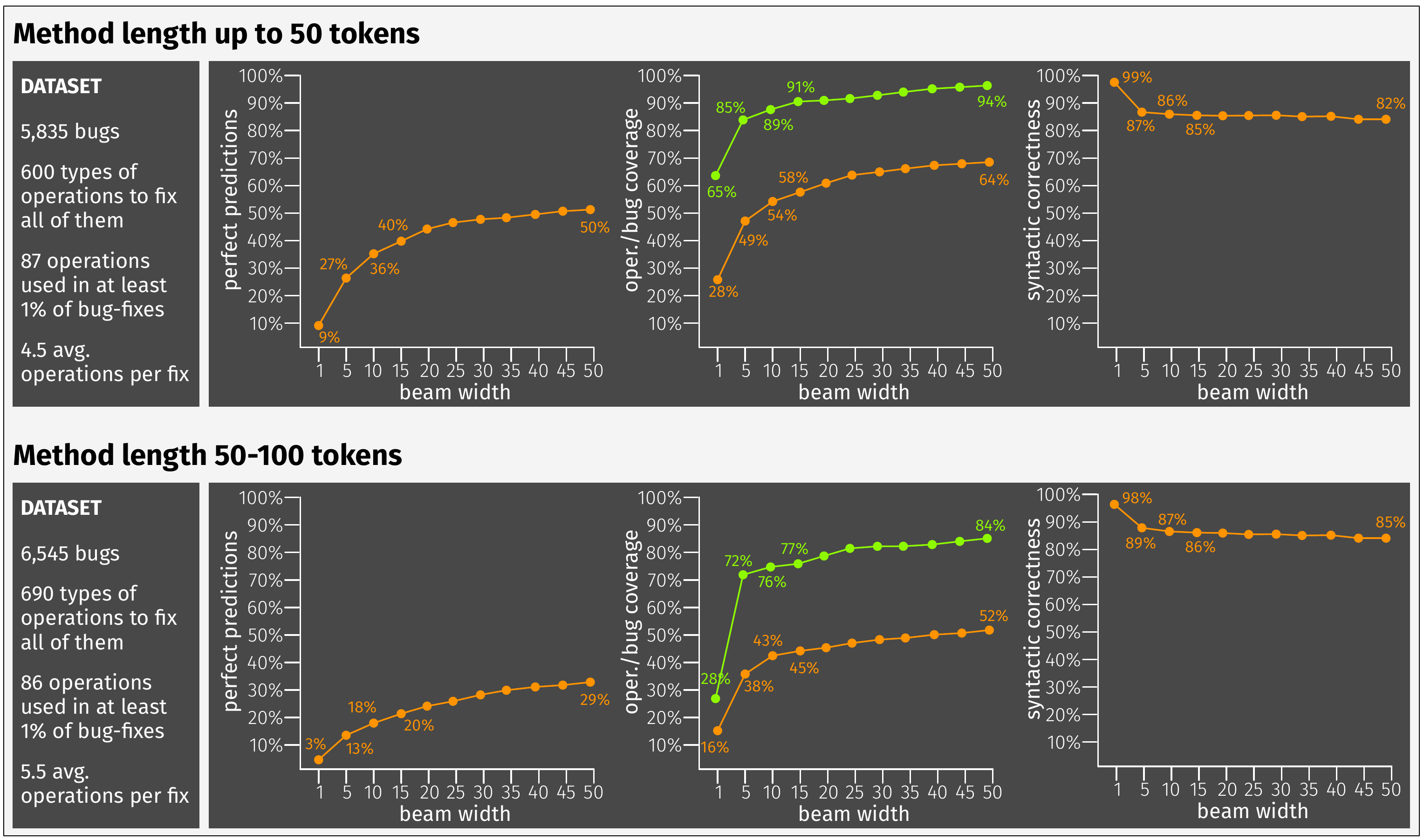}
	\caption{Number of perfect prediction, operation (orange) bug (green) coverage, and syntactic correctness for varying beam width and for different method lengths.}
	\label{fig:results}
\end{figure*}

\subsection{RQ1: Is Neural Machine Translation a viable approach to learn how to fix code?}
When performing the hyperparameter search, we found that the configuration, which achieved the best results on the validation set, for both $M_{small}$ and $M_{medium}$,  was the one with 1-layer bi-directional Encoder, 2-layer Attention Decoder both with 256 units, embedding size of 512, and  LSTM \cite{Hochreiter:1997:LSM:1246443.1246450} RNN cells. We trained the $M_{small}$ and $M_{medium}$ models for 50k and 55k epochs, respectively.

\subsubsection{Internal Dataset}
Table \ref{tab:rq1} reports the number and percentage of BFPs correctly predicted by the models for different beam sizes. As expected, increasing the beam size and, therefore, generating more candidate patches, increases the percentages of BFPs for which the models can perfectly generate the corresponding fixed code starting from the buggy code input. The most surprising results are those obtained with small beam sizes. The models can predict the fixed code of 9\% and 3\% of the BFPs with only one attempt. If we let the models generate 15 candidate patches, the percentage of perfect predictions bumps to 40\% and 20\% for small and medium methods, respectively. The number of BFPs patched steadily increases when more candidate patches are generated by the models (\ie bigger beam size), to reach a 50\% and 28\% of perfect predictions when 50 candidates patches are considered. 

The leftmost graphs in Fig. \ref{fig:results} show the percentage of successful fixes as a function of the beam size. When setting the beam size to 50, $M_{small}$ fixes 2,927 bugs (out of 5,835) in the same exact way they were fixed by developers. Likewise, $M_{medium}$ fixes 1,869 bugs (out of 6,545). It is important to note that all BFPs in the test sets are unique and have never been seen before by the model during the training or validation steps. Moreover, there is no inherent upper bound to the beam width used during inference, therefore even larger beam widths could be set. All perfect predictions generated by the models at different beam sizes as well as experiments with even larger beam sizes are available in our online appendix \cite{online-appendix}.

The differences in performances between the $M_{small}$ and $M_{medium}$ could be explained by the fact that larger methods have potentially more faulty locations where a transformation of the code could be performed. 

\subsubsection{External Dataset}
The last two columns of Table \ref{tab:rq1} report the results for the external dataset $CodRep$. The results show that the models $M_{small}$ and $M_{medium}$ -- which have been trained on the test set of $BFP_{small}$ and $BFP_{medium}$ -- are able to fix a high number of bugs belonging to a completely different and heterogeneous dataset. Similarly, increasing the beam size leads to improvements in the number of fixes, up to 26.66\% and 12.07\% for small and medium methods when 50 different potential patches are generated. Table \ref{tab:codrep_res} reports in details the fixes for each $CodRep$ dataset (beam size 50). The overall percentages of fixes for $CodRep$ dataset are slightly lower than those for $BFP$. This could be due to the fact that (i) the dataset is smaller, therefore, the model has fewer instances that can potentially fix; (ii) the dataset could contain different idioms which have not been considered (we did not fine-tune the idioms on this dataset). Overall, the results on the external dataset confirm the generalizability and potential of the NMT models to fix bugs on a different dataset.
\\
\begin{table}[t]
\scriptsize
\caption{Models' Performances}
\label{tab:rq1}
\centering
\begin{tabular}{l|c|c|c|c|}
\toprule
\multirow{ 2}{*}{Beam} & \multicolumn{2}{|c|}{$BFP$} & \multicolumn{2}{|c|}{$CodRep$}\\
 & $M_{small}$ & $M_{medium}$ & $M_{small}$ & $M_{medium}$\\
\midrule
1 & 538 / 5835 (\textbf{9.22}\%) & 211 / 6545 (\textbf{3.22}\%) & 65 / 3027 (\textbf{2.14}\%) & 26 / 6545 (\textbf{0.41}\%)\\
5 & 1595 / 5835 (\textbf{27.33}\%) & 859 / 6545 (\textbf{13.12}\%) & 311 / 3027 (\textbf{10.27}\%) & 207 / 6205 (\textbf{3.33}\%)\\
10 & 2119 / 5835 (\textbf{36.31}\%) & 1166 / 6545 (\textbf{17.82}\%) & 450 / 3027 (\textbf{14.86}\%) & 361 / 6205 (\textbf{5.81}\%)\\
15 & 2356 / 5835 (\textbf{40.37}\%) & 1326 / 6545 \textbf{(20.25}\%) & 539 / 3027 (\textbf{17.80}\%) & 451 / 6205 (\textbf{7.26}\%)\\
20 & 2538 / 5835 (\textbf{43.49}\%) & 1451 / 6545 (\textbf{22.16}\%) & 614 / 3027 (\textbf{20.28}\%) & 524 / 6205 (\textbf{8.44}\%)\\
25 & 2634 / 5835 (\textbf{45.14}\%) & 1558 / 6545 (\textbf{23.80}\%) & 657 / 3027 (\textbf{21.70}\%) & 574 / 6205 (\textbf{9.25}\%)\\
30 & 2711 / 5835 (\textbf{46.46}\%) & 1660 / 6545 (\textbf{25.36}\%) & 701 / 3027 (\textbf{23.15}\%) & 599 / 6205 (\textbf{9.65}\%)\\
35 & 2766 / 5835 (\textbf{47.40}\%) & 1720 / 6545 (\textbf{26.27}\%) & 733 / 3027 (\textbf{24.21}\%) & 644 / 6205 (\textbf{10.37}\%)\\
40 & 2834 / 5835 (\textbf{48.56}\%) & 1777 / 6545 (\textbf{27.15}\%) & 761 / 3027 (\textbf{25.14}\%) & 679 / 6205 (\textbf{10.94}\%)\\
45 & 2899 / 5835 (\textbf{49.68}\%) & 1830 / 6545 (\textbf{27.96}\%) & 784 / 3027 (\textbf{25.90}\%) & 709 / 6205 (\textbf{11.42}\%)\\
50 & 2927 / 5835 (\textbf{50.16}\%) & 1869 / 6545 (\textbf{28.55}\%) & 807 / 3027 (\textbf{26.66}\%) & 749 / 6205 (\textbf{12.07}\%)\\
\bottomrule
\end{tabular}
\end{table}

{\em \underline{\textbf{Summary for RQ$_{1}$}.}} 

Using NMT, we trained a model on small BFPs, which can produce developer inspired fixes for 9.22\% - 50.16\% of bugs (dependent upon beam width).  Likewise, a model trained on medium BFPs is capable of producing developer inspired fixes for 3.22\% - 28.55\% of bugs (dependent on beam width). These results indicate that Neural Machine Translation is a viable approach for learning how to fix code. The results on the external dataset confirm the generalizability and potential of the NMT models.

\begin{table}[t]
	\scriptsize
	\caption{$CodRep$ Results}
	\label{tab:codrep_res}
	\centering
	\begin{tabular}{l|c|c|}
		\toprule
		Datasets & $M_{small}$ & $M_{medium}$\\
		\midrule
		CodRep1 & 55 / 221 (\textbf{24.89}\%) & 48 / 503 (\textbf{9.54}\%)\\
		CodRep2 & 167 / 530 (\textbf{31.51}\%) & 136 / 1130 (\textbf{12.03}\%)\\
		CodRep3 & 180 / 665 (\textbf{27.07}\%) & 166 / 1397 (\textbf{11.88}\%)\\
		CodRep4 & 167 / 603 (\textbf{27.69}\%) & 157 / 1124 (\textbf{13.97}\%)\\
		CodRep5 & 288 / 1066 (\textbf{27.02}\%) & 244 / 2119 (\textbf{11.51}\%)\\
		\midrule
		$CodRep$ & 807 / 3027 (\textbf{26.66}\%) & 749 / 6205 (\textbf{12.07}\%)\\
		\bottomrule
	\end{tabular}
\end{table}

\subsection{RQ2: What types of operations are performed by the models?}

Fig. \ref{fig:results} also shows the results of the two models (\ie $M_{small}$ top, $M_{medium}$ bottom) in terms of operations coverage, and syntactic correctness of the generated patches. Before discussing these results, it is important to comment on the dataset characteristics for small and medium BFPs. To fix the 5,835 small methods, developers adopted combinations of 600 different types of operations at the AST level (\eg Insert BinaryOperator at Conditional, Delete Catch at Try, \etc). Of these, only 87 have been used in more than 1\% of bug-fixes, meaning that a vast majority of the AST operations have been rarely used to fix bugs (\eg in the case of the $BFP_{small}$, 513 types of AST operations have been used for the fixing of less than 58 bugs). Also, the average number of operations needed to fix a bug in the ``small'' dataset is 4.5. Similar observations can be done for $BFP_{medium}$ (see Fig. \ref{fig:results}).

\begin{table}[t]
	\scriptsize
	\caption{Perfect Prediction Top-10 Operations}
	\label{tab:top_operations}
	\centering
	\begin{tabular}{rl}
		\toprule
		\multicolumn{2}{c}{\textit{Delete}}\\
		\midrule
		Frequency & Operation\\
		\midrule
		1774 & Delete Invocation at Block\\
		1295 & Delete TypeAccess at ThisAccess\\
		1034 & Delete FieldRead at Invocation\\
		878  & Delete VariableRead at Invocation\\
		850  & Delete TypeAccess at Invocation\\
		789  & Delete Literal at Invocation\\
		622  & Delete TypeAccess at FieldRead\\
		602  & Delete ThisAccess at FieldRead\\
		570  & Delete ThisAccess at Invocation\\
		532  & Delete Literal at BinaryOperator\\
		\midrule
		\multicolumn{2}{c}{\textit{Insert}}\\
		\midrule
		Frequency & Operation\\
		\midrule
		338 & Insert TypeAccess at ThisAccess\\
		186 & Insert Literal at BinaryOperator\\
		185 & Insert Block at Method\\
		156 & Insert ThisAccess at FieldRead\\
		155 & Insert BinaryOperator at If\\
		150 & Insert ThisAccess at Invocation\\
		147 & Insert If at Block\\
		136 & Insert VariableRead at Invocation\\
		123 & Insert Return at Block\\
		114 & Insert Block at If\\
		\midrule
		\multicolumn{2}{c}{\textit{Move}}\\
		\midrule
		Frequency & Operation\\
		\midrule
		277 & Move Invocation from Block to CtInvocationImpl\\
		129 & Move Block from If to CtBlockImpl\\
		97  & Move Return from Block to CtReturnImpl\\
		76  & Move Invocation from BinaryOperator to CtInvocationImpl\\
		71  & Move Invocation from Invocation to CtInvocationImpl\\
		59  & Move Parameter from Method to CtParameterImpl\\
		59  & Move BinaryOperator from BinaryOperator to CtBinaryOperatorImpl\\
		57  & Move Invocation from LocalVariable to CtInvocationImpl\\
		57  & Move Block from Method to CtBlockImpl\\
		40  & Move Method from Class to CtMethodImpl\\
		\midrule
		\multicolumn{2}{c}{\textit{Update}}\\
		\midrule
		Frequency & Operation\\	
		\midrule
	    314 & Update Invocation at Block\\
		303 & Update Method at Class\\
		122 & Update Invocation at Invocation\\
		114 & Update Literal at Invocation\\
		77  & Update BinaryOperator at If\\
		75  & Update Invocation at Return\\
		48  & Update Literal at BinaryOperator\\
		45  & Update BinaryOperator at BinaryOperator\\
		38  & Update Invocation at LocalVariable\\
		33  & Update ThisAccess at FieldRead\\
		\bottomrule
	\end{tabular}
\end{table}

\subsubsection{Syntactic Correctness}
We start by analyzing the syntactic correctness (rightmost graphs). We can notice that, when the models are asked to generate a single prediction (\ie the most likely one), the overall syntactic correctness of the predicted code is very high (99\% and 98\%). Clearly, the more candidate predictions the model is required to generate, the more likely is that it introduces syntactic errors during the transformation of the code. We observe this phenomenon in the graph with a decreasing syntactic correctness, reaching 82\% and 85\% when 50 variants of patches are generated. The slightly better syntactic correctness achieved by the $M_{medium}$ model could be explained by the fact that, in larger methods, there are more potential points of transformation where syntactically correct variants can be generated, with respect to smaller methods. While we do not measure the compilability rate of the generated patches, it is worth to note that the perfect predictions generated by the models correspond to the code that was actually committed to repositories by developers. For such reasons, we could reasonably expect those predicted patches to be compilable.

\subsubsection{AST Operations}
The center graphs in Fig. \ref{fig:results} show the operation coverage (orange line) and theoretical bug coverage (green line) when varying the beam size. When only one candidate patch is generated, the models $M_{small}$ and $M_{medium}$ cover 28\% and 16\% of the total unique operations in the entire test sets, which include 600 and 690 operations, respectively. An increase of the beam size to 5 and 10 leads to a dramatic surge in the coverage of various operations in the test set. These results show that allowing the models to generate more candidate patches not only leads to more fixes, but also to a larger variety of bug fixing operations being performed. The operation coverage keeps increasing with larger beam widths. 

We observe a similar trend for the theoretical bug coverage, with large improvements in the early beam widths, and a steady increase afterwards. It is also worth to carefully speculate on the theoretical bug coverage percentages. As a matter of fact, the results suggest that -- with combinations of the AST actions learned and successfully emulated by the models in perfect fixes -- the models could theoretically cover 94\% and 84\% of the bug fixes in the test set. This means that the AST operations that the models failed to learn are quite rare, and only used in a small subset of the real bug-fixing activities used for training.

\tabref{tab:top_operations} reports the top-10 operations for each category successfully emulated by the models when generating the perfect predictions. The complete list of all the AST operations as extracted by GumTree Diff is available in our online appendix \cite{online-appendix}

\begin{figure*}[t!]
	\centering
	\includegraphics[width=\linewidth]{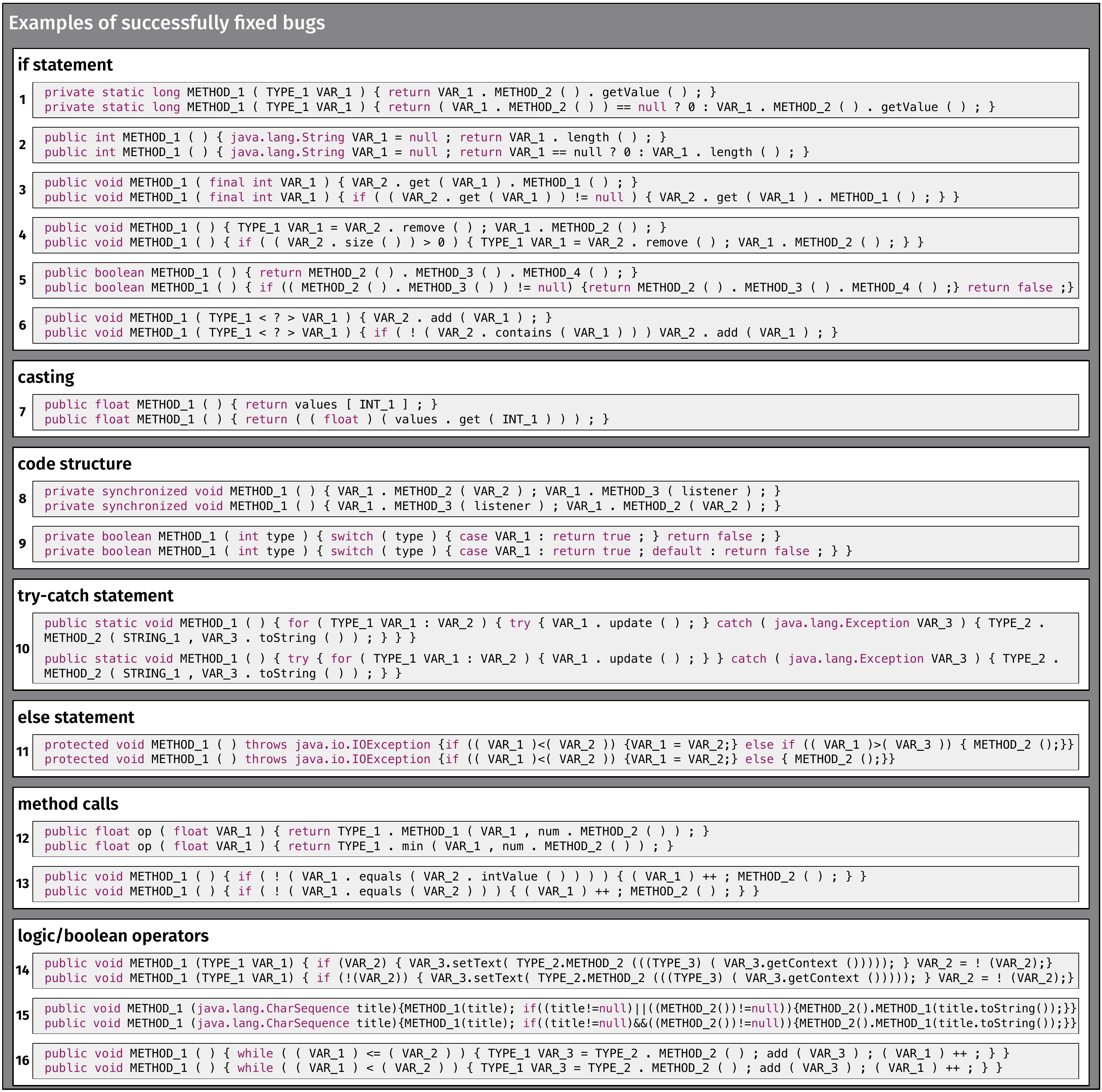}
	\caption{Examples of successfully-generated patches.}
	\label{fig:qualitative}
\end{figure*}

\subsubsection{Qualitative Examples}

Fig. \ref{fig:qualitative} shows some interesting examples of patches generated by the model. For space limitations, we focus on interesting fixes distilled from the set of perfect predictions generated by the model $M_{small}$. The examples are shown in abstracted code (with idioms), as they are fed and generated by the models. The actual source code can be generated by mapping back all the IDs to the real values stored in the mapping $M$. Fig. \ref{fig:qualitative} also groups the examples based on the ``type of fix'' implemented, showing the ability of the model in learning different fixing patterns, also in the context of the same group. For example, we show that not all fixes dealing with $\mathtt{if}$ conditions are identical. All examples are perfect predictions meaning that the model changed the buggy method to reflect exactly how the developer changed the method in the wild. 

Our first group of examples (1-6) concern buggy methods that were missing and $\mathtt{if}$ or that benefited from its addition. Thus, the added condition either helped to prevent errors during execution or ensured the expected outcome from the executed code. Example 1 shows an in-line \texttt{if} condition added to the fixed method to check whether the \texttt{getValue} method is being called on a null object and returns 0 if it is. If the object is not null, then the original \texttt{getValue} method is called on the object and that value is returned. This fix ensures that the get method is not called on a null object. Likewise, example 2 inserts a similar check but targeting the length of a variable rather than a getter method. The in-line \texttt{if} checks ensures the variable is not null, if it is then the method returns 0, otherwise, the method returns the result of \texttt{VAR\_1.length()}. This patch is interesting because the string \texttt{VAR\_1} is set to null before the length is calculated. This could be an example where the developer was performing error checking and wanted to ensure that the string was in fact nulled out. Given the context, we would expect the method to always return 0. However, the original method calls the \texttt{length()} method on a null string, which would return an error. Although this patch is ``correct'', the model may have learned a sub-optimal coding practice from the developer's patch. The model's generated patch is unsurprising as the training data included many in-line if checks as a potential fix. Most likely, the model identified this pattern and applied a similar fix in this scenario. Examples 3, 4 and 5 all insert similar if-checks that are not in-line \texttt{ifs}. Examples 3 and 5 both add an \texttt{if} condition handling cases in which the invoked method returns null, while example 4's \texttt{if} condition checks the size of a variable before operating on it. 

The last example in this group (\ie number 6) is different from the others, since the \texttt{if} condition is more complex and makes use of the boolean operator not (!). Here the fix is preventing the method from adding a duplicated value to \texttt{VAR\_2}. If the value \texttt{VAR\_1} is already present in \texttt{VAR\_2}, then the method will not add \texttt{VAR\_1} again. It is important to note that although these examples all add an \texttt{if} check as the fix, they are all unique and tailored to the method's context. The model was able to learn the correct changes needed for the specific method that would mimic a developer's changes.   

The second group of fixes addresses issues related to the cast of a specific variable. The original method in example 7 would throw an error if executed because the method signature calls for a return value of type float, but the method returns a value of type int. The model recognized this error and casted the return value of type float. Additionally, the fix also changes the mechanism by which the value is extracted from \texttt{values} (see Fig. \ref{fig:qualitative}). This not only changes the type of value returned by also the mechanism by which it is returned. 

The next group of examples pertain to the implementation or structure of the code, leading to incorrect execution. Example 8 switches the statements' order of execution, without applying any other change. This swap could be needed due to the first statement changing the state of the system (\eg the value of \texttt{VAR\_1}) which would then cause the \texttt{VAR\_1.METHOD\_2(VAR\_2)} invocation to have a different outcome. Our model is capable of finding such errors in order execution and provide an adequate fix. Example 9 is similar in that the structure of the code is incorrect. Here the \texttt{switch} statement is missing a \texttt{default} case in the buggy method. Thus, the buggy method will execute the \texttt{switch} and, if no \texttt{case} condition will be met, the code outside the \texttt{switch} statement will be run. The fix adds a default case to the \texttt{switch} statement to handle cases in which no \texttt{case} condition is met. This fix does not change the outcome of the code since the code executed outside the \texttt{case} statement (buggy version) and inside the \texttt{default} statement (fixed version) is exactly the same (see Fig. \ref{fig:qualitative}). However, it improves the readability of the code, making it adhering to the Java coding convention suggesting that \texttt{switch} statements should have a default case, which occurs when no other case in the \texttt{switch} has been met. 

Our fourth category of examples are changes where the model fixes \texttt{try-catch} statements.
We report one representative example (number 10). This fix changes the scope of the \texttt{try} block to also include in it the \texttt{for} loop, that was instead containing the \texttt{try} block in the buggy method.

The fifth group of fixes we found addresses incorrect \texttt{else} statements. In example 11 we see that the \texttt{else~if} statement is removed from the buggy method. This change is seen as a bug fix since the buggy method only defines its behavior when \texttt{VAR\_1 < VAR\_2} or \texttt{VAR\_1 > VAR\_2}. It has no behavior defined when \texttt{VAR\_1 == VAR\_2}, which could lead to unexpected errors. The model fixes this by replacing the \texttt{else if} statement with an \texttt{else}, covering all possible relations between \texttt{VAR\_1} and \texttt{VAR\_2}.

The sixth group of fixes aims at replacing incorrect method calls. As seen in example 12, the method call \texttt{METHOD\_1} is replaced with \texttt{min}. The example demonstrates the power of idioms. Indeed, without this idiom, we would discard this fix since we would be unable to generate a name for the unseen method \texttt{min} in the fix and would name it \texttt{METHOD\_3}. Since \texttt{METHOD\_3} would not be seen in our mapping $M$, we would have to synthesize the new methods name when translating the abstracted code back into source code. Having \texttt{min} as an idiom allows us to avoid the synthesis and still learn the fix. Although this patch replaces \texttt{METHOD\_1} with \texttt{min}, which seems very specific, it is logical that the method is being replaced with a mathematical method. The power of our RNN is that context is taken into account and the context of this method demonstrates a mathematical operation. When the model observes mathematical methods it may be inclined to generate a patch using variables and method calls seen in similar methods. Therefore, the \texttt{min} function makes sense as a patch since it compares two numbers and is a mathematical function, which fits the context of the overall method well.  Example 13 shows instead the removal of unnecessary/harmful method calls. Here the model removes in the fixed method the invocation to \texttt{intValue()} on \texttt{VAR\_2}. This method is used to return a numeric value, represented by an object, as an \texttt{int}. In this situation, texttt{VAR\_2} is a Java integer object and \texttt{intValue()} would return an \texttt{int} type. The fix removes this method call which compares an object to \texttt{int}, making \texttt{equals} comparing \texttt{VAR\_1} to the integer type \texttt{VAR\_2}.

Finally, the last group of fixes involves the changing, addition or removal of logic or boolean operators. Although they changes themselves do not appear massive, they have major implications on the source code behavior. For instance, example 14 adds a negation boolean operator to the \texttt{if} condition. This completely changes the functionality of the fixed method since now it will only execute the \texttt{if} block when \texttt{! VAR\_2 == true}. Example 15 performs a fix along the same line, changing an operator in the \texttt{if} condition from logical or (\textbar\textbar) to a logical and (\&\&). This means that both conditions must be met in order to run the code within the \texttt{if} block. Since the buggy method allowed only one condition to be met, it is possible that this led to undesired results for the developers. Example 16 changes a $<=$ to a $<$ operator. It is worth noting that this operator change takes place within the scope of a \texttt{while} loop, thus reducing by one the times that the code in the \texttt{while} loop is executed.

The reported qualitative examples show the potential of NMT models to generate meaningful correct patches, by learning from real bug-fixes wrote by developers, which allows the model to avoid problems arising with existing program repair techniques. Indeed, a previous work by Qi \etal \cite{Qi:2015} found that existing techniques achieve repair  by overfitting on the test cases, or by simply deleting pieces of functionality. The models produced many other interesting patches, which are not discussed here due to space limitations. Our online appendix \cite{online-appendix} contains many more examples of bug-fixes using different operations,  considering methods with different lengths, and using a variety of beam widths.

\begin{figure*}[t!]
	\centering
	\includegraphics[width=\linewidth]{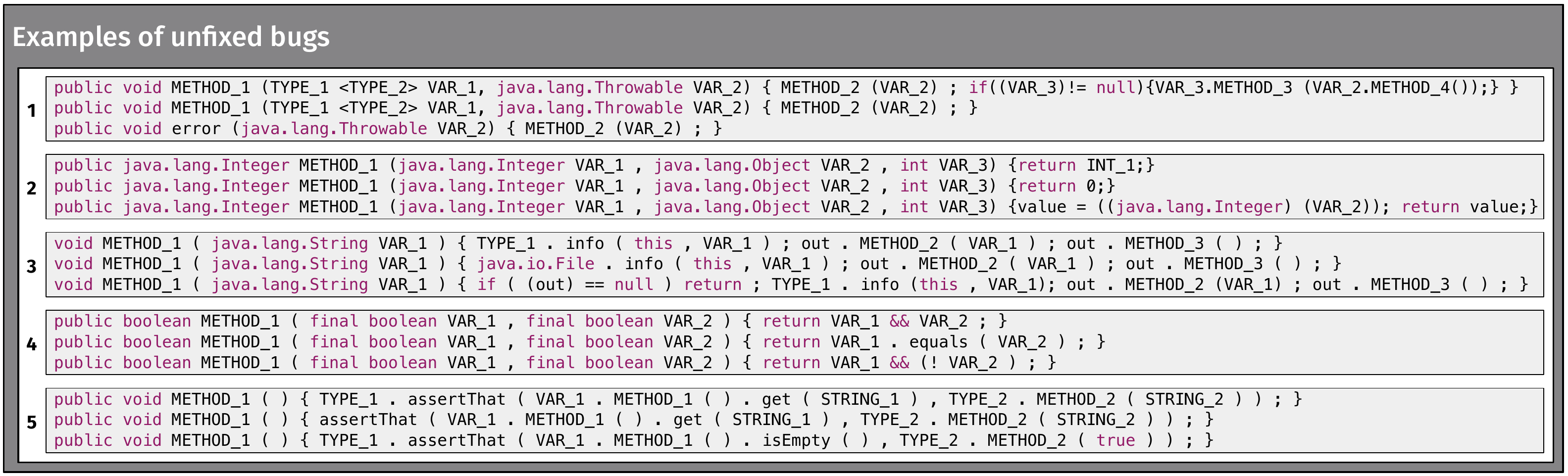}
	\caption{Examples of unfixed bugs.}
	\label{fig:qualitative-negative}
\end{figure*}

Besides success cases, we also qualitatively discuss code snippets for which our approach failed in fixing the bug. Each of the discussed examples is shown in \figref{fig:qualitative-negative} in three lines: the first represent the buggy code to fix, the second is the fix generated by our approach, and the third is the correct fix implemented by the developer.

In the first example, the model deleted the last {\tt if} statement from the buggy code. However, the developer fix aimed at renaming the method, removing \texttt{VAR\_1} from the method parameters and deleting the {\tt if}-block at the end of the method body. These three changes were not completely captured by the model.  In fact, the only fix that was implemented was the removal of the if-block at the end of the method body. Most likely, the model failed to completely fix this bug due to the drastic changes to the original method and a lack of training data to properly identify this pattern of changes. From a practical standpoint, the renaming of the method is a refactoring operation rather than a bug fix, and the model's fix would have the same functionality as the developer fix. Although the model's fix is not perfect, the functionality is correct.

In the second example, the model's fix changes the return statement from an integer of unknown value, to the idiom \texttt{0}. However, the developer's fix is much more extensive as it assigns to \texttt{value} the integer cast of \texttt{VAR\_2}. After this integer cast of \texttt{VAR\_2} has been performed, \texttt{value} is returned. In this example, the developer has introduced a completely new statement and a variable as a bug fix. The model struggles with synthesizing new statements unless the pattern is repeatedly seen in the training data. It is likely that this pattern of changes was not seen frequently during training and, therefore, the model performed the fix incorrectly.

The third example demonstrates the model's inability to recognize the need for an {\tt if}-block in the beginning of the method body. The model attempted to fix the bug by changing the type of \texttt{TYPE\_1 . info ( this , VAR\_1 )}. The developer's fix shows that the type of that variable is correct. However, an {\tt if}-check was missing to determine whether the {\tt out} variable was null. Without this check, methods could be called on a null object, which would lead to an error. The model has previously inserted these {\tt if-checks} in other examples; however, in this particular case, the model failed to recognize the potential for a null object.

The fourth example shows the model incorrectly changing the return value. The developer's performed fix on this method was to add the Boolean \texttt{not operator: !} to \texttt{VAR\_2}. Our model changed the and operator \texttt{\&\&} to a \texttt{. equals ( )} method call.  The model's fix is the opposite of the developer's fix:  the developer's method will return true whereas the model's method would return false, and \emph{vice versa}. One explanation for the model's behavior could be the frequent case of changing \texttt{==} to \texttt{. equals ( )}, which we see often in the training data. Although the \texttt{\&\&} symbol is not the same as \texttt{==}, the same symbol twice in a row pattern may cause the model to replace that symbol with \texttt{. equals ( )}.

Finally, the fifth example shows that the model struggles to find various, drastic changes that were implemented by the developer. The model's ``fix'' in this situation is to remove the type from the {\tt assertThat} method call. However, the developer has fixed this method by replacing the \texttt{METHOD\_1.get(STRING\_1)} with \texttt{METHOD\_1 . isEmpty ( )} and then replacing the \texttt{STRING\_2} argument in \texttt{TYPE\_2 . METHOD\_2 ( STRING\_2 )} with the idiom \texttt{true}. The model fails to recognize both changes and therefore would be ineffective in finding a bug fix for the original method. The developer's method completely changes the computation and the purpose from the original method, which may be why the model is not capable to perform a fix.

{\em \underline{\textbf{Summary for RQ$_{2}$}.}} The models exhibit a very high syntactic correctness of the generated patches ranging between 99\% and 82\%. Moreover, while the models are able to learn on how to apply a subset of the AST operation types exploited by developers to fix all bugs in the test set, the learned operations are the most representative ones, allowing to, theoretically, fix a large percentage of bugs.

\begin{figure}[h]
	\centering
	\includegraphics[width=0.4\linewidth]{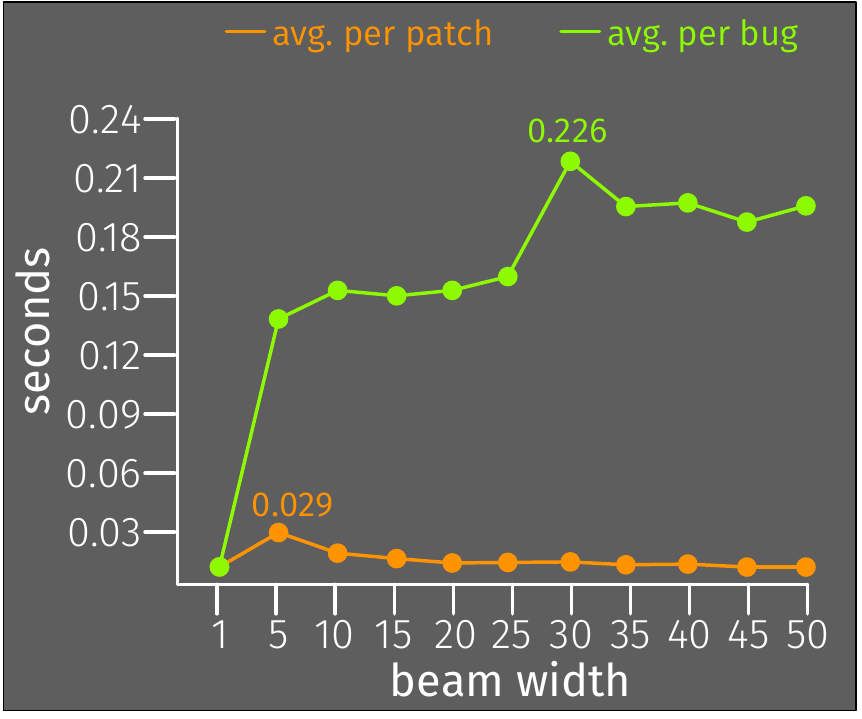}
	\caption{Inference Time ($M_{medium}$).}
	\label{fig:time}
\end{figure}

\subsection{RQ3: What is the training and inference time of the models?}
The training of the models  $M_{small}$ and $M_{medium}$ took six and 15 hours respectively, running on a server with three consumer-level GPUs. 
Overall, this is an acceptable one-time cost that allows building a cross-project bug-fixing model in a reasonable amount of time. Fig. \ref{fig:time} shows the average inference time per patch (orange line) and per bug (green line) for the $M_{medium}$ model with increasingly large beam size. While the average time per bug rises with larger beam sizes (\ie more patches generated for the same bug) from a minimum of only 0.006s ($k=1$) to a maximum of 0.226s ($k=30$), the average time per patch generated stays well below 0.030s. Overall, the model is able to generate 50 candidate patches for a bug in less than a second. The inference times for $M_{small}$ are even lower. The complete timing results, raw values, and total number of seconds are available in our online appendix \cite{online-appendix}.

{\em \underline{\textbf{Summary for RQ$_{3}$}.}} After training for less than 15 hours, the models are able to generate 50 candidate patches for a single bug in less than a second.

	\section{Threats to Validity}
	\label{sec:threats}

\textbf{Construct validity} threats concern the relationship between theory and observation, and are mainly related to likely sources of imprecision in our analyses. To have enough training data, we mined bug-fixes in GitHub repositories rather than using curated bug-fix datasets such as Defects4j \cite{Just:ISSTA14} or IntroClass\cite{DBLP:journals/tse/GouesHSBDFW15}, useful but very limited in size. To mitigate imprecisions in our datasets, we manually analyzed a sample of the extracted commits and verified that they were related to bug-fixes.

\textbf{Internal validity} threats concern factors internal to our study that could influence our results.   It is possible that the performance of our models depends on the hyperparameter configuration. We explain in \secref{sec:hyperpar} how hyperparameter search has been performed.

\textbf{External validity} threat concern the generalizability of our findings. We did not compare NMT models with state-of-the-art techniques supporting automatic program repair since our main goal was not to propose a novel approach for automated program repair, but rather to execute a large-scale empirical study investigating the suitability of NMT for generating patches. Additional steps are needed to convert the methodology we adopted into an end-to-end working tool, such as the automatic implementation of the patch, or the execution of the test cases for checking a patch's suitability. This is part of our future work agenda.

We only focused on Java programs. However, the learning process is language-independent and the whole infrastructure can be instantiated for different programming languages by replacing the lexer, parser and AST differencing tools.

Finally, we only focused on small- and medium-sized methods. We reached this decision after analyzing the distribution of the extracted BFPs, balancing the amount of training data available and the variability in sentence length.

	\section{Related Work}
	\label{sec:related}
This section describes related work on (i) automated program repair techniques and, specifically, their underlying redundancy assumption, and (ii) the use of machine translation to support software engineering tasks.

\subsection{Program~Repair and~the~Redundancy~Assumption}
\label{sub:automated}

Automated program repair involves the transformation of an unacceptable behavior of a program execution into an acceptable one according to a specification~\cite{Monperrus:2015}. Behavioral repair techniques in particular change the behavior of a program under repair by changing its source or binary code~\cite{Monperrus:2015}. These techniques~\cite{DBLP:journals/tse/GouesNFW12,LeGoues:2012,Sidiroglou-Douskos:2015} rely on a critical assumption, the \emph{redundancy assumption}, that claims large programs contain the seeds of their own repair. This assumption has been examined by at least two independent empirical studies, showing that a significant proportion of commits originates from previously-existing code~\cite{Martinez:2014,Barr:2014}.

Martinez \etal~\cite{Martinez:2014} empirically examined the assumption that certain bugs can be fixed by copying and rearranging existing code. They validated the redundancy assumption by defining a concept of \emph{software temporal redundancy}. A commit is temporally redundant if it is a rearrangement of code in previous commits. They measured redundancy at two levels of granularity: line- and token-level.  At line-level granularity, they found that most of the temporal redundancy is localized in the same file. At token-level granularity, their results imply that many repairs never need to invent a new token.

Barr \etal~\cite{Barr:2014} examined a history of 15,723 commits to determine the extent to which the commits can be reconstructed from existing code. The grafts they found were mostly single lines, \ie micro-clones, and they proposed that micro-clones are useful since they are the atoms of code construction~\cite{Barr:2014}. Their findings align with Martinez \etal~\cite{Martinez:2014} in that changes to a codebase contain fragments that already exist in the code base at the time of the change.

Repair approaches based on the redundancy assumption are called redundancy-based repair techniques, since they leverage redundancy and repetition in source code~\cite{Pierret:2009,Gabel:2010,Hindle:2012,Carzaniga:2013,Nguyen:2013:ASE,Martinez:2014,Barr:2014,White:2019:SANER}.
For example, GenProg~\cite{DBLP:journals/tse/GouesNFW12,LeGoues:2012,LeGoues:2012:GECCO} searches for statement-level modifications to make to an abstract syntax tree. The approach by Arcuri and Yao \cite{DBLP:conf/cec/ArcuriY08} co-evolves programs and test cases using a model similar to the predatory-prey one. Weimer \etal \cite{Weimer:2013:LPE:3107656.3107702} perform program repair using a deterministic search, reducing the search space with program equivalence analysis.
Le \etal \cite{DBLP:conf/wcre/LeLG16} use the content of previous patches to reward patches that can have a likely better acceptability for developers, and therefore avoid over-fitting patches to test cases. 
A complementary set of repair techniques leverage program analysis and program synthesis {to repair programs by constructing code with particular properties~\cite{Nguyen:2013:ICSE,Mechtaev:2015,Ke:2015,Xuan:2016,Mechtaev:2016, Tian:2017:ADR:3106237.3106300, Le:2017:SSS:3106237.3106309,Perkins:2009:APE:1629575.1629585}.

Some approaches perform automated program repair by searching the fix among manually-written patterns, as in the case of 
PAR \cite{DBLP:conf/icse/KimNSK13}, or through a SMT (satisfiability modulo theories)-based semantic code search over a large body of existing code, as for SearchRepair \cite{Ke:2015}.
Instead, Prophet~\cite{Long:2016:APG:2837614.2837617} is a learning-based approach that uses explicitly designed code features to rank candidate repairs. Other approaches train on correct solutions (from student programs) to specific programming tasks and try to learn task-specific repair strategies~\cite{Bhatia:2016,Pu:2016}. This goal has been achieved successfully in contexts such as in massively open online courses (MOOC), where the programs are generally small and synthetic~\cite{Gupta:2017}.

Finally, one of the first approaches able to learn code transformation templates from human-written patches is Genesis, by Long \etal \cite{Long:2017:AIC:3106237.3106253}. Genesis can learn patches related to three types of defects, specifically related to null pointers, out of bounds and class casting. Long \etal show how Genesis is able to find more patches than previously-proposed approaches like PAR \cite{DBLP:conf/icse/KimNSK13}. Also, the approach proposed by Long \etal is capable to infer new tokens from existing patches.
Our approach is different from the work of Long \etal, as we employ NMT whereas they use integer programming. Also, differently from Genesis, in our work we would like to experiment with the applicability of NMT to learn and then apply generic (and not defect-specific) fixes from a large body of source code changes.

The goal of our empirical investigation was to determine whether NMT could be used to bring the ``redundancy assumption'', but also the heuristics used by program repair approaches using code search, at a next level. Such a next level would be the capability to automatically learn patches from large software corpora.

As mentioned in the introduction, this work represent an extension of our previous work in which we proposed the general idea of learning bug-fixes using NMT \cite{Tufano:2018:EIL:3238147.3240732}. While our previous paper mainly presented the idea and assessed its overall feasibility, this paper reports an extensive evaluation, in which we also (i) generate multiple candidate patches via beam search; (ii) analyze the types of AST operations performed in the fixes as well as the syntactic correctness; (iii) qualitatively analyze the kinds of fix operations the learned models are able to perform, and (vi) assess the timing performance of the approach when learning the models and when recommending the fix.

\subsection{{Machine~Translation~in~Software~Engineering}}

Modern machine translation systems generally use data-driven methods to translate text or speech from one language to another. Machine translation systems are trained on translated texts, or ``parallel corpora'', for particular text types~\cite{Koehn:2010} in both natural languages and formal languages such as programming languages. Manually migrating software projects from one language to another is a time-consuming and error-prone task~\cite{Zhong:2010}. Nguyen \etal~\cite{Nguyen:2014:ASE,Nguyen:2013:FSE:2,Nguyen:2014} used \emph{statistical} machine translation for method-to-method migration from Java to C\#, translating small token sequences at a time. Recently, NMT systems superseded traditional statistical approaches as the state-of-the-art in translation. One advantage neural systems have over purely statistical systems is they can measure fluency at a higher level of granularity, \eg sentence-level granularity, rather than being constrained to phrases. However, NMT systems are indeed data-hungry systems, and this problem has been an issue for software engineering applications where there are not a lot of parallel corpora. DeepAM~\cite{Gu:2017} uses deep learning to automatically mine application programming interface mappings from a source code corpus without parallel text. Our work is intended to study the feasibility of using NMT to learn bug-fixes from real-world changes.
	
	\section{Future Work}
	\label{sec:future_work}

Our initial goal is to continue to improve and analyze the model we have generated. One task we plan to begin with is to analyze the trade off between our training set percentage, model quality and training time. We plan to use a tiered approach to determine how many training instances are needed to achieve an optimal performance. Therefore, we will train the model with 10\% of the training set, then increase this percentage to 20\%, 30\%, etc. all the way up to 90\%. This analysis will include a discussion about diminishing returns as well as the timing required to train the model. We currently don't implement this techinique in our work due to the extensive amount of time needed for the study. Training deep neural networks can take many days and is heavily reliant on the computational power and size of the dataset. As our dataset is fairly large, training on 10\% of the training set all the way up to 90\% will take a substantial amount of time. Then after the model is trained, we would still need to perform the analysis to determine the quality of the model. Given that this is a very expensive analysis, we save this task for future work.

As extensive future work, we want to focus on two major limitations of the approach. The first is to address a different level of granularity as opposed to function level. Currently, our approach only learns patches at a function level with less than 100 tokens. We would like to increase this size and maybe even change the granularity to class or package level. This change in granularity would increase the context for certain bug fixes and should make the approach more meaningful to developers. It may even be possible to generate bug fixes that span across multiple files, requiring a change to multiple methods.

Our second task is similar to the first, but we want to focus on learning smaller changes within the context of a larger method. We want to focus on a segmentation technique, which would allow us to keep the context meaningful but still allow the model to change tokens. Our idea is to further abstract the source code so that only meaningful statements are recognized. This technique can be difficult since further abstraction leads to less meaningful context. However, we believe that walking the line between less meaningful context and not overwhelming the model, is the key to success. We do not want to lose fixes we have found by changing types and variables within the code, therefore, we plan to combine our tool with state of the art static analysis tools. These latter should be able to find meaningful fixes pertaining to types and variables, which frees up our deep learning model to focus on more complex fixes. Thus, we plan to severely abstract any statement or code fragment that can be changed with a static analysis tool and feed the remaining parts of the large method into our deep learning RNN model. This methodology still maintains meaningful context while allowing the model to focus on more complex parts of the method, which static analysis may not be useful for.
	
	\section{Conclusion}
	\label{sec:conclusion}

We presented an extensive empirical investigation into the applicability of Neuro-Machine Translation (NMT) for the purpose of learning how to fix code, from real bug-fixes. We first devised and detailed a process to mine, extract, and abstract the source code of bug-fixes available in the wild, in order to obtain method-level examples of bug-fix pairs (BFPs). Then, we set up, trained, and tuned NMT models to \textit{translate} buggy code into fixed code. Our empirical analysis aimed at assessing the feasibility of the NMT technique applied to the bug-fixing problem, the types and quality of the predicted patches, as well as the training and inference time of the models.

We found the models to be able to fix a large number of unique bug-fixes, ranging between 9-50\% of small BFPs (up to 2,927 unique fixed bugs) and 3-28\% of medium BFPs (up to 1,869 unique fixed bugs) in our test set, depending on the amount of candidate patches we require the model to generate. The models generate syntactically correct patches in more than 82\% of the cases. The model $M_{small}$ is able to emulate between 28-64\% of the Abstract Syntax Tree operations performed during fixes, while $M_{medium}$ achieves between 16-52\% of the coverage. Finally, the running time analysis shows that these models are capable of generating tens of candidate patches in a split of a second.

This study constitutes a solid empirical foundation upon which other researchers could build, and appropriately evaluate, program repair techniques based on NMT.
	
	\section{Acknowledgment}
	\label{sec:acknowledgment}
We thank Zimin Chen, Steve Kommrusch, and Martin Monperrus for the valuable help in extracting and providing the $CodRep$ dataset. This work is supported in part by the NSF CCF-1525902 and CCF-1815186 grants. Gabriele Bavota gratefully acknowledges the financial support of the Swiss National Science Foundation for the CCQR project (SNF Project No. 175513). Any opinions, findings, and conclusions expressed herein are the authors and do not necessarily reflect those of the sponsors. 
	
	\bibliographystyle{acm}
	\bibliography{ms}

\begin{thebibliography}{10}

\bibitem{DBLP:conf/iwpc/AlaliKM08}
{\sc Alali, A., Kagdi, H.~H., and Maletic, J.~I.}
\newblock What's a typical commit? {A} characterization of open source software
  repositories.
\newblock In {\em The 16th {IEEE} International Conference on Program
  Comprehension, {ICPC} 2008, Amsterdam, The Netherlands, June 10-13, 2008\/}
  (2008), pp.~182--191.

\bibitem{Allamanis:duplication}
{\sc Allamanis, M.}
\newblock The adverse effects of code duplication in machine learning models of
  code.

\bibitem{Allamanis:2015:SAM:2786805.2786849}
{\sc Allamanis, M., Barr, E.~T., Bird, C., and Sutton, C.}
\newblock Suggesting accurate method and class names.
\newblock In {\em Proceedings of the 2015 10th Joint Meeting on Foundations of
  Software Engineering\/} (New York, NY, USA, 2015), ESEC/FSE 2015, ACM,
  pp.~38--49.

\bibitem{AntoniolAPKG08}
{\sc Antoniol, G., Ayari, K., {Di Penta}, M., Khomh, F., and
  Gu{\'{e}}h{\'{e}}neuc, Y.}
\newblock Is it a bug or an enhancement?: a text-based approach to classify
  change requests.
\newblock In {\em Proceedings of the 2008 conference of the Centre for Advanced
  Studies on Collaborative Research, October 27-30, 2008, Richmond Hill,
  Ontario, Canada\/} (2008), p.~23.

\bibitem{DBLP:conf/cec/ArcuriY08}
{\sc Arcuri, A., and Yao, X.}
\newblock A novel co-evolutionary approach to automatic software bug fixing.
\newblock In {\em Proceedings of the {IEEE} Congress on Evolutionary
  Computation, {CEC} 2008, June 1-6, 2008, Hong Kong, China\/} (2008),
  pp.~162--168.

\bibitem{DBLP:journals/corr/BahdanauCB14}
{\sc Bahdanau, D., Cho, K., and Bengio, Y.}
\newblock Neural machine translation by jointly learning to align and
  translate.
\newblock {\em CoRR abs/1409.0473\/} (2014).

\bibitem{Barr:2014}
{\sc Barr, E.~T., Brun, Y., Devanbu, P., Harman, M., and Sarro, F.}
\newblock The plastic surgery hypothesis.
\newblock In {\em Proceedings of the 22Nd ACM SIGSOFT International Symposium
  on Foundations of Software Engineering\/} (New York, NY, USA, 2014), FSE
  2014, ACM, pp.~306--317.

\bibitem{Bhatia:2016}
{\sc Bhatia, S., and Singh, R.}
\newblock Automated correction for syntax errors in programming assignments
  using recurrent neural networks.
\newblock {\em CoRR abs/1603.06129\/} (2016).

\bibitem{boulanger2013audio}
{\sc Boulanger-Lewandowski, N., Bengio, Y., and Vincent, P.}
\newblock Audio chord recognition with recurrent neural networks.
\newblock In {\em ISMIR\/} (2013), Citeseer, pp.~335--340.

\bibitem{DBLP:journals/corr/BritzGLL17}
{\sc Britz, D., Goldie, A., Luong, M., and Le, Q.~V.}
\newblock Massive exploration of neural machine translation architectures.
\newblock {\em CoRR abs/1703.03906\/} (2017).

\bibitem{Brown:2017:CFW:3106237.3106280}
{\sc Brown, D.~B., Vaughn, M., Liblit, B., and Reps, T.}
\newblock The care and feeding of wild-caught mutants.
\newblock In {\em Proceedings of the 2017 11th Joint Meeting on Foundations of
  Software Engineering\/} (New York, NY, USA, 2017), ESEC/FSE 2017, ACM,
  pp.~511--522.

\bibitem{Carzaniga:2013}
{\sc Carzaniga, A., Gorla, A., Mattavelli, A., Perino, N., and Pezz\`{e}, M.}
\newblock Automatic recovery from runtime failures.
\newblock In {\em Proceedings of the 2013 International Conference on Software
  Engineering\/} (Piscataway, NJ, USA, 2013), ICSE '13, IEEE Press,
  pp.~782--791.

\bibitem{codrep}
{\sc Chen, Z., and Monperrus, M.}
\newblock {CodRep}.
\newblock \url{https://github.com/KTH/CodRep-competition}, 2018.

\bibitem{arXiv-1807.03200}
{\sc Chen, Z., and Monperrus, M.}
\newblock The codrep machine learning on source code competition.
\newblock Tech. Rep. 1807.03200, arXiv, 2018.

\bibitem{DBLP:journals/corr/ChoMGBSB14}
{\sc Cho, K., van Merrienboer, B., G{\"{u}}l{\c{c}}ehre, {\c{C}}., Bougares,
  F., Schwenk, H., and Bengio, Y.}
\newblock Learning phrase representations using {RNN} encoder-decoder for
  statistical machine translation.
\newblock {\em CoRR abs/1406.1078\/} (2014).

\bibitem{DBLP:conf/kbse/FalleriMBMM14}
{\sc Falleri, J., Morandat, F., Blanc, X., Martinez, M., and Monperrus, M.}
\newblock Fine-grained and accurate source code differencing.
\newblock In {\em {ACM/IEEE} International Conference on Automated Software
  Engineering, {ASE} '14, Vasteras, Sweden - September 15 - 19, 2014\/} (2014),
  pp.~313--324.

\bibitem{DBLP:conf/icsm/FischerPG03}
{\sc Fischer, M., Pinzger, M., and Gall, H.~C.}
\newblock Populating a release history database from version control and bug
  tracking systems.
\newblock In {\em 19th International Conference on Software Maintenance {(ICSM}
  2003), The Architecture of Existing Systems, 22-26 September 2003, Amsterdam,
  The Netherlands\/} (2003), p.~23.

\bibitem{Gabel:2010}
{\sc Gabel, M., and Su, Z.}
\newblock A study of the uniqueness of source code.
\newblock In {\em Proceedings of the Eighteenth ACM SIGSOFT International
  Symposium on Foundations of Software Engineering\/} (New York, NY, USA,
  2010), FSE '10, ACM, pp.~147--156.

\bibitem{github-compare}
{\sc GitHub}.
\newblock {GitHub Compare API}.
\newblock
  \url{https://developer.github.com/v3/repos/commits/\#compare-two-commits},
  2010.

\bibitem{LeGoues:2012}
{\sc Goues, C.~L., Dewey-Vogt, M., Forrest, S., and Weimer, W.}
\newblock A systematic study of automated program repair: Fixing 55 out of 105
  bugs for \$8 each.
\newblock ICSE'12.

\bibitem{LeGoues:2012:GECCO}
{\sc Goues, C.~L., Weimer, W., and Forrest, S.}
\newblock Representations and operators for improving evolutionary software
  repair.
\newblock GECCO'12.

\bibitem{DBLP:journals/corr/abs-1211-3711}
{\sc Graves, A.}
\newblock Sequence transduction with recurrent neural networks.
\newblock {\em CoRR abs/1211.3711\/} (2012).

\bibitem{githubarchive}
{\sc Grigorik, I.}
\newblock {GitHub Archive}.
\newblock \url{https://www.githubarchive.org}, 2012.

\bibitem{DeepCodeSearch}
{\sc Gu, X., Zhang, H., and Kim, S.}
\newblock Deep code search.
\newblock In {\em Proceedings of the 40th International Conference on Software
  Engineering, {ICSE} 2018, Gothenburg, Sweden, May 27 - June 3, 2018\/}
  (2018).

\bibitem{DBLP:conf/sigsoft/GuZZK16}
{\sc Gu, X., Zhang, H., Zhang, D., and Kim, S.}
\newblock Deep {API} learning.
\newblock In {\em Proceedings of the 24th {ACM} {SIGSOFT} International
  Symposium on Foundations of Software Engineering, {FSE} 2016, Seattle, WA,
  USA, November 13-18, 2016\/} (2016), pp.~631--642.

\bibitem{Gu:2017}
{\sc Gu, X., Zhang, H., Zhang, D., and Kim, S.}
\newblock {DeepAM:} migrate {APIs} with multi-modal sequence to sequence
  learning.
\newblock {\em CoRR abs/1704.07734\/} (2017).

\bibitem{Gupta:2017}
{\sc Gupta, R., Kanade, A., and Shevade, S.~K.}
\newblock Deep reinforcement learning for programming language correction.
\newblock {\em CoRR abs/1801.10467\/} (2018).

\bibitem{ICSE.2012.6227193}
{\sc Hata, H., Mizuno, O., and Kikuno, T.}
\newblock Bug prediction based on fine-grained module histories.
\newblock {\em Proceedings - International Conference on Software
  Engineering\/} (06 2012), 200--210.

\bibitem{HerzigJZ13}
{\sc Herzig, K., Just, S., and Zeller, A.}
\newblock It's not a bug, it's a feature: how misclassification impacts bug
  prediction.
\newblock In {\em 35th International Conference on Software Engineering, {ICSE}
  '13, San Francisco, CA, USA, May 18-26, 2013\/} (2013), pp.~392--401.

\bibitem{Hindle:2012}
{\sc Hindle, A., Barr, E.~T., Su, Z., Gabel, M., and Devanbu, P.}
\newblock On the naturalness of software.
\newblock In {\em Proceedings of the 34th International Conference on Software
  Engineering\/} (Piscataway, NJ, USA, 2012), ICSE '12, IEEE Press,
  pp.~837--847.

\bibitem{Hochreiter:1997:LSM:1246443.1246450}
{\sc Hochreiter, S., and Schmidhuber, J.}
\newblock Long short-term memory.
\newblock {\em Neural Comput. 9}, 8 (Nov. 1997), 1735--1780.

\bibitem{Jin:2011:AAF:1993498.1993544}
{\sc Jin, G., Song, L., Zhang, W., Lu, S., and Liblit, B.}
\newblock Automated atomicity-violation fixing.
\newblock In {\em Proceedings of the 32Nd ACM SIGPLAN Conference on Programming
  Language Design and Implementation\/} (New York, NY, USA, 2011), PLDI '11,
  ACM, pp.~389--400.

\bibitem{Jorgensen:2007:SRS:1248721.1248736}
{\sc Jorgensen, M., and Shepperd, M.}
\newblock A systematic review of software development cost estimation studies.
\newblock {\em IEEE Trans. Softw. Eng. 33}, 1 (Jan. 2007), 33--53.

\bibitem{Just:ISSTA14}
{\sc Just, R., Jalali, D., and Ernst, M.~D.}
\newblock {Defects4J}: A database of existing faults to enable controlled
  testing studies for {Java} programs.
\newblock In {\em Proceedings of the 2014 International Symposium on Software
  Testing and Analysis\/} (New York, NY, USA, 2014), ISSTA 2014, ACM,
  pp.~437--440.

\bibitem{kalchbrenner-blunsom:2013:EMNLP}
{\sc Kalchbrenner, N., and Blunsom, P.}
\newblock Recurrent continuous translation models.
\newblock In {\em Proceedings of the 2013 Conference on Empirical Methods in
  Natural Language Processing\/} (Seattle, Washington, USA, October 2013),
  Association for Computational Linguistics, pp.~1700--1709.

\bibitem{Ke:2015}
{\sc Ke, Y., Stolee, K., {Le~Goues}, C., and Brun, Y.}
\newblock Repairing programs with semantic code search.
\newblock ASE'15.

\bibitem{DBLP:conf/icse/KimNSK13}
{\sc Kim, D., Nam, J., Song, J., and Kim, S.}
\newblock Automatic patch generation learned from human-written patches.
\newblock In {\em 35th International Conference on Software Engineering, {ICSE}
  '13, San Francisco, CA, USA, May 18-26, 2013\/} (2013), pp.~802--811.

\bibitem{Koehn:2010}
{\sc Koehn, P.}
\newblock {\em Statistical Machine Translation}.
\newblock 2010.

\bibitem{10.1007/978-3-642-35843-2_6}
{\sc Kolassa, C., Riehle, D., and Salim, M.~A.}
\newblock A model of the commit size distribution of open source.
\newblock In {\em SOFSEM 2013: Theory and Practice of Computer Science\/}
  (Berlin, Heidelberg, 2013), P.~van Emde~Boas, F.~C.~A. Groen, G.~F. Italiano,
  J.~Nawrocki, and H.~Sack, Eds., Springer Berlin Heidelberg, pp.~52--66.

\bibitem{DBLP:conf/iwpc/LamNNN17}
{\sc Lam, A.~N., Nguyen, A.~T., Nguyen, H.~A., and Nguyen, T.~N.}
\newblock Bug localization with combination of deep learning and information
  retrieval.
\newblock In {\em Proceedings of the 25th International Conference on Program
  Comprehension, {ICPC} 2017, Buenos Aires, Argentina, May 22-23, 2017\/}
  (2017), pp.~218--229.

\bibitem{Le:2017:SSS:3106237.3106309}
{\sc Le, X., Chu, D., Lo, D., {Le Goues}, C., and Visser, W.}
\newblock S3: Syntax- and semantic-guided repair synthesis via programming by
  examples.
\newblock FSE'17.

\bibitem{DBLP:conf/wcre/LeLG16}
{\sc Le, X.~D., Lo, D., and {Le Goues}, C.}
\newblock History driven program repair.
\newblock In {\em {IEEE} 23rd International Conference on Software Analysis,
  Evolution, and Reengineering, {SANER} 2016, Suita, Osaka, Japan, March 14-18,
  2016 - Volume 1\/} (2016), pp.~213--224.

\bibitem{DBLP:conf/icse/GouesDFW12}
{\sc {Le Goues}, C., Dewey{-}Vogt, M., Forrest, S., and Weimer, W.}
\newblock A systematic study of automated program repair: Fixing 55 out of 105
  bugs for {\textdollar}8 each.
\newblock In {\em 34th International Conference on Software Engineering, {ICSE}
  2012, June 2-9, 2012, Zurich, Switzerland\/} (2012), pp.~3--13.

\bibitem{DBLP:journals/tse/GouesHSBDFW15}
{\sc {Le Goues}, C., Holtschulte, N., Smith, E., Brun, Y., Devanbu, P.,
  Forrest, S., and Weimer, W.}
\newblock The manybugs and introclass benchmarks for automated repair of {C}
  programs.
\newblock {\em TSE 41}, 12 (2015), 1236--1256.

\bibitem{DBLP:journals/tse/GouesNFW12}
{\sc {Le Goues}, C., Nguyen, T., Forrest, S., and Weimer, W.}
\newblock Genprog: {A} generic method for automatic software repair.
\newblock {\em {IEEE} Trans. Software Eng. 38}, 1 (2012), 54--72.

\bibitem{DBLP:journals/corr/Li0KBLT16}
{\sc Li, D., Li, L., Kim, D., Bissyand{\'{e}}, T.~F., Lo, D., and Traon, Y.~L.}
\newblock Watch out for this commit! {A} study of influential software changes.
\newblock {\em CoRR abs/1606.03266\/} (2016).

\bibitem{Long:2017:AIC:3106237.3106253}
{\sc Long, F., Amidon, P., and Rinard, M.}
\newblock Automatic inference of code transforms for patch generation.
\newblock In {\em Proceedings of the 2017 11th Joint Meeting on Foundations of
  Software Engineering\/} (New York, NY, USA, 2017), ESEC/FSE 2017, ACM,
  pp.~727--739.

\bibitem{Long:2016:APG:2837614.2837617}
{\sc Long, F., and Rinard, M.}
\newblock Automatic patch generation by learning correct code.
\newblock In {\em Proceedings of the 43rd Annual ACM SIGPLAN-SIGACT Symposium
  on Principles of Programming Languages\/} (New York, NY, USA, 2016), POPL
  '16, ACM, pp.~298--312.

\bibitem{DBLP:journals/corr/LuongPM15}
{\sc Luong, M., Pham, H., and Manning, C.~D.}
\newblock Effective approaches to attention-based neural machine translation.
\newblock {\em CoRR abs/1508.04025\/} (2015).

\bibitem{DBLP:journals/ese/MartinezDSXM17}
{\sc Martinez, M., Durieux, T., Sommerard, R., Xuan, J., and Monperrus, M.}
\newblock Automatic repair of real bugs in {J}ava: a large-scale experiment on
  the {Defects4J} dataset.
\newblock {\em Empirical Software Engineering 22}, 4 (2017), 1936--1964.

\bibitem{Martinez:2014}
{\sc Martinez, M., Weimer, W., and Monperrus, M.}
\newblock Do the fix ingredients already exist? an empirical inquiry into the
  redundancy assumptions of program repair approaches.
\newblock In {\em Companion Proceedings of the 36th International Conference on
  Software Engineering\/} (New York, NY, USA, 2014), ICSE Companion 2014, ACM,
  pp.~492--495.

\bibitem{Mechtaev:2016}
{\sc Mechtaev, S., Jooyong, Y., and Roychoudhury, A.}
\newblock Angelix: Scalable multiline program patch synthesis via symbolic
  analysis.
\newblock ICSE'16.

\bibitem{Mechtaev:2015}
{\sc Mechtaev, S., Jooyong, Y., and Roychoudhury, A.}
\newblock {DirectFix}: Looking for simple program repairs.
\newblock ICSE'15.

\bibitem{Monperrus:2015}
{\sc Monperrus, M.}
\newblock Automatic software repair: A bibliography.
\newblock {\em ACM Comput. Surv. 51}, 1 (Jan. 2018), 17:1--17:24.

\bibitem{monperrus:hal-00769121}
{\sc Monperrus, M., and Martinez, M.}
\newblock {CVS-Vintage: A Dataset of 14 CVS Repositories of Java Software}.
\newblock working paper or preprint, Dec. 2012.

\bibitem{Moran:TSE2018}
{\sc Moran, K., Bernal-C\'{a}rdenas, C., Curcio, M., Bonett, R., and
  Poshyvanyk, D.}
\newblock Machine learning-based prototyping of graphical user interfaces for
  mobile apps.
\newblock {\em IEEE Transactions on Software Engineering\/} (2018).

\bibitem{Nguyen:2014:ASE}
{\sc Nguyen, A.~T., Nguyen, H.~A., Nguyen, T.~T., and Nguyen, T.~N.}
\newblock Statistical learning approach for mining {API} usage mappings for
  code migration.
\newblock In {\em Proceedings of the 29th ACM/IEEE International Conference on
  Automated Software Engineering\/} (New York, NY, USA, 2014), ASE '14, ACM,
  pp.~457--468.

\bibitem{Nguyen:2013:FSE:2}
{\sc Nguyen, A.~T., Nguyen, T.~T., and Nguyen, T.~N.}
\newblock Lexical statistical machine translation for language migration.
\newblock In {\em Proceedings of the 2013 9th Joint Meeting on Foundations of
  Software Engineering\/} (New York, NY, USA, 2013), ESEC/FSE 2013, ACM,
  pp.~651--654.

\bibitem{Nguyen:2014}
{\sc Nguyen, A.~T., Nguyen, T.~T., and Nguyen, T.~N.}
\newblock Migrating code with statistical machine translation.
\newblock In {\em Companion Proceedings of the 36th International Conference on
  Software Engineering\/} (New York, NY, USA, 2014), ICSE Companion 2014, ACM,
  pp.~544--547.

\bibitem{Nguyen:2013:ASE}
{\sc Nguyen, H.~A., Nguyen, A.~T., Nguyen, T.~T., Nguyen, T.~N., and Rajan, H.}
\newblock A study of repetitiveness of code changes in software evolution.
\newblock In {\em Proceedings of the 28th IEEE/ACM International Conference on
  Automated Software Engineering\/} (Piscataway, NJ, USA, 2013), ASE'13, IEEE
  Press, pp.~180--190.

\bibitem{Nguyen:2013:ICSE}
{\sc Nguyen, H. D.~T., Qi, D., Roychoudhury, A., and Chandra, S.}
\newblock Semfix: Program repair via semantic analysis.
\newblock In {\em Proceedings of the 2013 International Conference on Software
  Engineering\/} (Piscataway, NJ, USA, 2013), ICSE '13, IEEE Press,
  pp.~772--781.

\bibitem{Parr:2013:DAR:2501720}
{\sc Parr, T.}
\newblock {\em The Definitive {ANTLR} 4 Reference}, 2nd~ed.
\newblock Pragmatic Bookshelf, 2013.

\bibitem{Parr:2011:LFA:1993498.1993548}
{\sc Parr, T., and Fisher, K.}
\newblock Ll(*): The foundation of the {ANTLR} parser generator.
\newblock In {\em Proceedings of the 32Nd ACM SIGPLAN Conference on Programming
  Language Design and Implementation\/} (New York, NY, USA, 2011), PLDI '11,
  ACM, pp.~425--436.

\bibitem{Perkins:2009:APE:1629575.1629585}
{\sc Perkins, J.~H., Kim, S., Larsen, S., Amarasinghe, S., Bachrach, J.,
  Carbin, M., Pacheco, C., Sherwood, F., Sidiroglou, S., Sullivan, G., Wong,
  W.-F., Zibin, Y., Ernst, M.~D., and Rinard, M.}
\newblock Automatically patching errors in deployed software.
\newblock In {\em Proceedings of the ACM SIGOPS 22Nd Symposium on Operating
  Systems Principles\/} (New York, NY, USA, 2009), SOSP '09, ACM, pp.~87--102.

\bibitem{Pierret:2009}
{\sc Pierret, D., and Poshyvanyk, D.}
\newblock An empirical exploration of regularities in open-source software
  lexicons.
\newblock In {\em The 17th {IEEE} International Conference on Program
  Comprehension, {ICPC} 2009, Vancouver, British Columbia, Canada, May 17-19,
  2009\/} (2009), pp.~228--232.

\bibitem{Pu:2016}
{\sc Pu, Y., Narasimhan, K., Solar-Lezama, A., and Barzilay, R.}
\newblock Sk\_p: A neural program corrector for {MOOC}s.
\newblock SPLASH Companion 2016.

\bibitem{Qi:2015}
{\sc Qi, Z., Long, F., Achour, S., and Rinard, M.}
\newblock An analysis of patch plausibility and correctness for
  generate-and-validate patch generation systems.
\newblock ISSTA'15.

\bibitem{Raychev:2014:CCS:2594291.2594321}
{\sc Raychev, V., Vechev, M., and Yahav, E.}
\newblock Code completion with statistical language models.
\newblock In {\em Proceedings of the 35th ACM SIGPLAN Conference on Programming
  Language Design and Implementation\/} (New York, NY, USA, 2014), PLDI '14,
  ACM, pp.~419--428.

\bibitem{DBLP:conf/iwpc/RoyC08a}
{\sc Roy, C.~K., and Cordy, J.~R.}
\newblock {NICAD:} accurate detection of near-miss intentional clones using
  flexible pretty-printing and code normalization.
\newblock In {\em The 16th {IEEE} International Conference on Program
  Comprehension, {ICPC} 2008, Amsterdam, The Netherlands, June 10-13, 2008\/}
  (2008), pp.~172--181.

\bibitem{Sajnani:2016:SSC:2884781.2884877}
{\sc Sajnani, H., Saini, V., Svajlenko, J., Roy, C.~K., and Lopes, C.~V.}
\newblock {SourcererCC}: Scaling code clone detection to big-code.
\newblock In {\em Proceedings of the 38th International Conference on Software
  Engineering\/} (New York, NY, USA, 2016), ICSE '16, ACM, pp.~1157--1168.

\bibitem{Scholtes2016}
{\sc Scholtes, I., Mavrodiev, P., and Schweitzer, F.}
\newblock From aristotle to ringelmann: a large-scale analysis of team
  productivity and coordination in open source software projects.
\newblock {\em Empirical Software Engineering 21}, 2 (Apr 2016), 642--683.

\bibitem{seacord:2003:MLS:599767}
{\sc seacord, R.~C., Plakosh, D., and Lewis, G.~A.}
\newblock {\em Modernizing Legacy Systems: Software Technologies, Engineering
  Process and Business Practices}.
\newblock Addison-Wesley Longman Publishing Co., Inc., Boston, MA, USA, 2003.

\bibitem{Sidiroglou-Douskos:2015}
{\sc Sidiroglou-Douskos, S., Lahtinen, E., Long, F., and Rinard, M.}
\newblock Automatic error elimination by horizontal code transfer across
  multiple applications.
\newblock {\em SIGPLAN Not. 50}, 6 (June 2015), 43--54.

\bibitem{Smith:2015:CWD:2786805.2786825}
{\sc Smith, E.~K., Barr, E.~T., Le~Goues, C., and Brun, Y.}
\newblock Is the cure worse than the disease? overfitting in automated program
  repair.
\newblock In {\em Proceedings of the 2015 10th Joint Meeting on Foundations of
  Software Engineering\/} (New York, NY, USA, 2015), ESEC/FSE 2015, ACM,
  pp.~532--543.

\bibitem{DBLP:conf/wcre/SobreiraDDMM18}
{\sc Sobreira, V., Durieux, T., Delfim, F.~M., Monperrus, M., and {de Almeida
  Maia}, M.}
\newblock Dissection of a bug dataset: Anatomy of 395 patches from defects4j.
\newblock In {\em 25th International Conference on Software Analysis, Evolution
  and Reengineering, {SANER} 2018, Campobasso, Italy, March 20-23, 2018\/}
  (2018), pp.~130--140.

\bibitem{soto2018using}
{\sc Soto, M., and Le~Goues, C.}
\newblock Using a probabilistic model to predict bug fixes.
\newblock In {\em 2018 IEEE 25th International Conference on Software Analysis,
  Evolution and Reengineering (SANER)\/} (2018), IEEE, pp.~221--231.

\bibitem{DBLP:journals/corr/SutskeverVL14}
{\sc Sutskever, I., Vinyals, O., and Le, Q.~V.}
\newblock Sequence to sequence learning with neural networks.
\newblock {\em CoRR abs/1409.3215\/} (2014).

\bibitem{Tian:2017:ADR:3106237.3106300}
{\sc Tian, Y., and Ray, B.}
\newblock Automatically diagnosing and repairing error handling bugs in {C}.
\newblock In {\em Proceedings of the 2017 11th Joint Meeting on Foundations of
  Software Engineering\/} (New York, NY, USA, 2017), ESEC/FSE 2017, ACM,
  pp.~752--762.

\bibitem{doi:10.1002/smr.1797}
{\sc Tufano, M., Bavota, G., Poshyvanyk, D., Di~Penta, M., Oliveto, R., and
  De~Lucia, A.}
\newblock An empirical study on developer-related factors characterizing
  fix-inducing commits.
\newblock {\em Journal of Software: Evolution and Process 29}, 1, e1797.
\newblock e1797 JSME-15-0185.R2.

\bibitem{Tufano:2019:Changes}
{\sc Tufano, M., Pantiuchina, J., Watson, C., Bavota, G., and Poshyvanyk, D.}
\newblock On learning meaningful code changes via neural machine translation.
\newblock In {\em Proceedings of the 41st {ACM/IEEE} International Conference
  on Software Engineering\/} (2019), ICSE 19, ACM.

\bibitem{Tufano:2018:Similarities}
{\sc Tufano, M., Watson, C., Bavota, G., Di~Penta, M., White, M., and
  Poshyvanyk, D.}
\newblock Deep learning similarities from different representations of source
  code.
\newblock In {\em Proceedings of the 15th International Conference on Mining
  Software Repositories\/} (New York, NY, USA, 2018), MSR '18, ACM,
  pp.~542--553.

\bibitem{Tufano:2018:EIL:3238147.3240732}
{\sc Tufano, M., Watson, C., Bavota, G., Di~Penta, M., White, M., and
  Poshyvanyk, D.}
\newblock An empirical investigation into learning bug-fixing patches in the
  wild via neural machine translation.
\newblock In {\em Proceedings of the 33rd ACM/IEEE International Conference on
  Automated Software Engineering\/} (New York, NY, USA, 2018), ASE 2018, ACM,
  pp.~832--837.

\bibitem{online-appendix}
{\sc Tufano, M., Watson, C., Bavota, G., Di~Penta, M., White, M., and
  Poshyvanyk, D.}
\newblock {Online Appendix}.
\newblock \url{https://sites.google.com/view/learning-fixes}, 2018.

\bibitem{javaparser}
{\sc {van Bruggen}, D.}
\newblock {J}ava{P}arser.
\newblock \url{https://javaparser.org/about.html}, 2014.

\bibitem{WangLT16}
{\sc Wang, S., Liu, T., and Tan, L.}
\newblock Automatically learning semantic features for defect prediction.
\newblock In {\em Proceedings of the 38th International Conference on Software
  Engineering, {ICSE} 2016, Austin, TX, USA, May 14-22, 2016\/} (2016),
  pp.~297--308.

\bibitem{Weimer:2013:LPE:3107656.3107702}
{\sc Weimer, W., Fry, Z.~P., and Forrest, S.}
\newblock Leveraging program equivalence for adaptive program repair: Models
  and first results.
\newblock In {\em Proceedings of the 28th IEEE/ACM International Conference on
  Automated Software Engineering\/} (Piscataway, NJ, USA, 2013), ASE'13, IEEE
  Press, pp.~356--366.

\bibitem{DBLP:conf/icse/WeimerNGF09}
{\sc Weimer, W., Nguyen, T., {Le Goues}, C., and Forrest, S.}
\newblock Automatically finding patches using genetic programming.
\newblock In {\em 31st International Conference on Software Engineering, {ICSE}
  2009, May 16-24, 2009, Vancouver, Canada, Proceedings\/} (2009),
  pp.~364--374.

\bibitem{Weiss:2017:TIS:3155562.3155641}
{\sc Weiss, A., Guha, A., and Brun, Y.}
\newblock Tortoise: Interactive system configuration repair.
\newblock In {\em Proceedings of the 32nd IEEE/ACM International Conference on
  Automated Software Engineering\/} (Piscataway, NJ, USA, 2017), ASE 2017, IEEE
  Press, pp.~625--636.

\bibitem{Weiss:2007:LTF:1268983.1269017}
{\sc Weiss, C., Premraj, R., Zimmermann, T., and Zeller, A.}
\newblock How long will it take to fix this bug?
\newblock In {\em Proceedings of the Fourth International Workshop on Mining
  Software Repositories\/} (Washington, DC, USA, 2007), MSR '07, IEEE Computer
  Society, pp.~1--.

\bibitem{White:2019:SANER}
{\sc White, M., Tufano, M., Martinez, M., Monperrus, M., and Poshyvanyk, D.}
\newblock Sorting and transforming program repair ingredients via deep learning
  code similarities.
\newblock In {\em 2019 IEEE 26th International Conference on Software Analysis,
  Evolution and Reengineering (SANER)\/} (2019), IEEE, p.~to appear.

\bibitem{WhiteTVP16}
{\sc White, M., Tufano, M., Vendome, C., and Poshyvanyk, D.}
\newblock Deep learning code fragments for code clone detection.
\newblock In {\em Proceedings of the 31st {IEEE/ACM} International Conference
  on Automated Software Engineering, {ASE} 2016, Singapore, September 3-7,
  2016\/} (2016), pp.~87--98.

\bibitem{White:2015}
{\sc White, M., Vendome, C., {Linares-V{\'a}squez}, M., and Poshyvanyk, D.}
\newblock Toward deep learning software repositories.
\newblock MSR'15.

\bibitem{Xuan:2016}
{\sc Xuan, J., Mart{\'i}nez, M., DeMarco, F., Cl{\'e}ment, M., Lamelas, S.,
  Durieux, T., {Le~Berre}, D., and Monperrus, M.}
\newblock {Nopol}: Automatic repair of conditional statement bugs in {Java}
  programs.
\newblock {\em IEEE Trans. Software Engineering 43}, 1 (2016), 34--55.

\bibitem{Yang:2017:BTC:3106237.3106274}
{\sc Yang, J., Zhikhartsev, A., Liu, Y., and Tan, L.}
\newblock Better test cases for better automated program repair.
\newblock In {\em Proceedings of the 2017 11th Joint Meeting on Foundations of
  Software Engineering\/} (New York, NY, USA, 2017), ESEC/FSE 2017, ACM,
  pp.~831--841.

\bibitem{Zhong:2015:ESR:2818754.2818864}
{\sc Zhong, H., and Su, Z.}
\newblock An empirical study on real bug fixes.
\newblock In {\em Proceedings of the 37th International Conference on Software
  Engineering - Volume 1\/} (Piscataway, NJ, USA, 2015), ICSE '15, IEEE Press,
  pp.~913--923.

\bibitem{Zhong:2010}
{\sc Zhong, H., Thummalapenta, S., Xie, T., Zhang, L., and Wang, Q.}
\newblock Mining {API} mapping for language migration.
\newblock In {\em Proceedings of the 32Nd ACM/IEEE International Conference on
  Software Engineering - Volume 1\/} (New York, NY, USA, 2010), ICSE '10, ACM,
  pp.~195--204.

\bibitem{Zhou:2012:BFM:2337223.2337226}
{\sc Zhou, J., Zhang, H., and Lo, D.}
\newblock Where should the bugs be fixed? - more accurate information
  retrieval-based bug localization based on bug reports.
\newblock In {\em Proceedings of the 34th International Conference on Software
  Engineering\/} (Piscataway, NJ, USA, 2012), ICSE '12, IEEE Press, pp.~14--24.

\end{thebibliography}

\end{document}